\documentclass[twocolumn]{aastex63}

\usepackage{threeparttable}
\usepackage{booktabs}
\usepackage{soul}

\usepackage{enumitem}
\setlist[itemize]{leftmargin=*}
\usepackage{amsmath}

\shorttitle{X-ray Variability of Seyferts -- I. Rms Spectral Diversity}
\shortauthors{Hu et al.}

\begin{document}

\title{A Systematic Study of the Short-Term X-ray Variability of Seyfert Galaxies \\
I. Diversity of the X-ray Rms Spectra}

\correspondingauthor{Jingwei Hu, Chichuan Jin}
\email{hujingwei@nao.cas.cn, ccjin@nao.cas.cn}

\author{Jingwei Hu}
\affiliation{Key Laboratory of Space Astronomy and Technology, National Astronomical Observatories, Chinese Academy of Sciences, \\
Beijing, 100101, People's Republic of China}
\affiliation{School of Astronomy and Space Sciences, University of Chinese Academy of Sciences, 19A Yuquan Road, Beijing, 100049, \\
People's Republic of China}

\author{Chichuan Jin}
\affiliation{Key Laboratory of Space Astronomy and Technology, National Astronomical Observatories, Chinese Academy of Sciences, \\
Beijing, 100101, People's Republic of China}
\affiliation{School of Astronomy and Space Sciences, University of Chinese Academy of Sciences, 19A Yuquan Road, Beijing, 100049, \\
People's Republic of China}

\author{Huaqing Cheng}
\affiliation{Key Laboratory of Space Astronomy and Technology, National Astronomical Observatories, Chinese Academy of Sciences, \\
Beijing, 100101, People's Republic of China}

\author{Weimin Yuan}
\affiliation{Key Laboratory of Space Astronomy and Technology, National Astronomical Observatories, Chinese Academy of Sciences, \\
Beijing, 100101, People's Republic of China}
\affiliation{School of Astronomy and Space Sciences, University of Chinese Academy of Sciences, 19A Yuquan Road, Beijing, 100049, \\
People's Republic of China}

\begin{abstract}
The X-ray variability of active galactic nuclei (AGN) carries crucial information about the X-ray radiation mechanism. We performed a systematic study of the X-ray short-term (1--100 ks timescale) variability for a large sample of 78 Seyferts with 426 deep {\it XMM-Newton} observations. In this paper, we present the time-averaged spectra and rms spectra for the entire sample, which show a variety of properties. Based on the spectral shape, we divide the rms spectra into five subtypes and the time-averaged spectra into four subtypes. The most common shape of the rms spectra is concave-down where the rms peaks at $\sim$ 1 keV. We find that different sources can show similar time-averaged spectra and rms spectra. However, there is no one-to-one mapping between the subtypes of the time-averaged spectra and rms spectra, as similar time-averaged spectra can be accompanied by different rms spectra, and vice versa. This is likely because different physical mechanisms can produce similar rms spectra. For every subtype of the time-averaged spectra, we report its preferred subtypes of the rms spectra in both low- and high-frequency bands. We also compare the statistical properties for different subtypes, such as the black hole mass and Eddington ratio. Finally, we investigate the rms in the Fe K$\alpha$ line regime and find that those with a broad and extended red-wing profile tend to show stronger variability than those showing a narrow or relatively symmetric profile. Our results demonstrate the necessity of performing joint spectral and variability modeling in order to understand the mechanism of the X-ray emission in AGN. All of the rms spectra have been made publicly available.
\end{abstract}


\section{Introduction}
\label{Introduction}
Active galactic nuclei (AGN) are powered by the accretion of materials onto supermassive black holes (SMBHs) lying in the center of galaxies. AGN are powerful radiation sources in the universe, whose X-ray emission is believed to originate from high-energy processes in the vicinity of SMBH, such as the inverse Compton scattering in a high-temperature optically thin corona \citep[e.g.][]{1991ApJ...380L..51H}. Other components or processes that complicate the X-ray emission include a separate warm corona \citep[e.g.][]{1997ApJ...477...93L,1998MNRAS.301..179M,2012MNRAS.420.1848D}, reflection by the accretion disk or distant materials \citep[e.g.][]{2004MNRAS.349.1435M,2005MNRAS.358..211R,2013MNRAS.429.2917F}, absorption by the partially ionized/neutral gas \citep[e.g.][]{2007A&A...475..121T,2007ARA&A..45..441M,2012ApJ...752...94T}, contribution from the galaxy (e.g. thermal plasma emission associated with star formation; \citealt{1989ARA&A..27...87F,2003A&A...399...39R}) and emission by the ionized gas on large scale (e.g. \citealt{2018MNRAS.481..947K,2021MNRAS.508.1798P}). In order to explore the X-ray mechanism of AGN, detailed modeling of their time-averaged spectra is often conducted. However, it is found that different models can provide similarly well fits to the same time-averaged spectrum and are thus degenerate. For instance, the origin of the soft X-ray excess is a matter of long debate, as it can be well fitted by both the ionized reflection model \citep[e.g.][]{2005MNRAS.358..211R,2006MNRAS.365.1067C,2013MNRAS.428.2795K,2013MNRAS.429.2917F} and the warm Comptonization model \citep[e.g.][]{1997ApJ...477...93L,1998MNRAS.301..179M,2012MNRAS.420.1848D}.

In order to break the model degeneracy and uncover the underlying physics of the X-ray emission, it is necessary to also take the X-ray variability into consideration. A commonly used quantity to measure the intensity of X-ray variability is the fractional root-mean-square (rms), $F_{\rm var}$, which is defined as the square root of the excess variance \citep{2002ApJ...568..610E,2003ApJ...598..935M,2003MNRAS.345.1271V}. In the frequency domain, the excess variance can be calculated by integrating the power spectral density over a range of frequencies with the Poisson noise power subtracted \citep{2003MNRAS.345.1271V, 2008MNRAS.389.1427P, 2014A&ARv..22...72U}. It is also possible to derive $F_{\rm var}$ for the light curves in different energy bands. Therefore, $F_{\rm var}$ can be measured both in a given energy band and for a given frequency range (e.q. timescale), and it is also model-independent. The frequency-differentiated rms spectra can complement the time-averaged spectral analysis and help distinguishing different spectral components \citep[e.g.][]{2011MNRAS.417..250M,2013MNRAS.436.3173J,2017MNRAS.468.3663J,2020MNRAS.492.1363P,2021MNRAS.500.2475J,2021MNRAS.500.4506H}.

Previous studies focusing on individual Seyferts have shown a variety of X-ray rms spectra. For example,
\citet{2002MNRAS.335L...1F} presented a long hard look at the Seyfert 1 galaxy MCG-6-30-15 with {\it XMM-Newton} and confirmed the presence of the broad asymmetric Fe line profile.
They found that the rms spectrum show less variability at low and high energy, and the Fe line is not as variable as the continuum.
This concave-down shape of the rms spectrum peaking at 1-2 keV was also observed in some other Seyferts \citep[e.g.][]{2004MNRAS.348.1415V,2007ApJ...656..116M},  which can be interpreted as the constant soft excess and Compton reflection components plus the variable power-law component \citep[e.g.][]{2003MNRAS.342L..31T}.
\citet{2011MNRAS.417..250M} performed a detailed variability analysis for RE J1034+396 and found the rms spectrum rising smoothly with energy (also see \citealt{2021MNRAS.500.2475J}). A similar shape of the rms spectrum was also found in the super-Eddington NLS1 RX J0136.9-3510 (\citealt{2009MNRAS.398L..16J}). 

The shape of the rms spectrum also depends on the variability timescale (i.e. frequency band; e.g. \citealt{2008MNRAS.387..279A,2009MNRAS.394..250M}).
\citet{2013MNRAS.436.3173J} presented a detailed spectral-timing analysis for PG 1244+026 using a 120 ks {\it XMM-Newton} observation.
They found at low frequencies ($8\times10^{-6} - 1\times10^{-4}$ Hz) that the soft X-rays were more variable than the hard X-rays, while the opposite was true at high frequencies ($2\times10^{-4} - 5\times10^{-3}$ Hz).
There may exist several X-ray components with different variability properties and dominating different energy bands. Alternatively, the variable component can change its spectral shape at different time scales.
\citet{2020A&A...634A..65D} reported on an X-ray spectral-timing analysis for NGC 3783 and detected a mildly decreasing and concave-down trend of the rms spectra in two epochs.
They suggested that the change in the rms spectrum is due to incoherent fast variability of X-ray obscuration on timescales between about 1 hour to 10 hours. \citet{2021MNRAS.508.1798P} reported the different rms spectra of the NLS1 1H 0707-495 in different frequency bands and modeled them with variability introduced by both reflection and absorption.

While previous works have shown that the rms spectrum is useful for studying the X-ray mechanism, they are mostly limited to individual sources. The main objective of this work is to obtain a systematic understanding of the diversity of the X-ray rms spectra of Seyferts, and look for potential connections between the rms spectra and time-averaged spectra. To achieve this objective, we collect a large sample of 78 Seyferts showing significant short-term X-ray variability (1-100 ks timescales) and use 426 archival {\it XMM-Newton} observations to measure the time-averaged spectra and rms spectra in different frequency bands for all the sources.

This paper is organized as follows. In Section~\ref{DataReduction}, we introduce the selection criteria and statistical properties of the sample and the data analysis procedures. In Section~\ref{sec-statistics}, we present a statistical view of the diversity of rms spectra for the entire sample. This is followed by Section~\ref{sec-connection}, which presents various systematic connections between the time-averaged spectra and rms spectra. The variability analysis of Fe K$\alpha$ is presented in Section~\ref{FeK}. Further discussions about the diverse rms spectra and the variation in the rms spectrum between different observations of individual sources are provided in Section~\ref{discussion}. The final section summarizes all the main results of this work.

\section{Sample Selection and Data Reduction} \label{DataReduction}
\subsection{Sample Selection} \label{sample}
We first adopt the sample of 43 Seyfert galaxies from \citet{2016MNRAS.462..511K}, in which the authors performed a global look at their X-ray time lags. The sample included Seyferts observed with the {\it XMM-Newton} \citep{2001A&A...365L...1J} up to 2015 January 1, with continuous observations longer than 40 ks.
By using the same selection criterion, Seyferts with observations publicly available between 2015 January 1 and 2020 January 1 are also added into our sample. Furthermore, for all the selected sources, we use all the {\it XMM-Newton} observations longer than 10 ks. This results in a large sample of 78 Seyfert galaxies with a total of 426 {\it XMM-Newton} observations.
We note that our Seyfert sample contains six radio-loud sources with 18 observations and is thus dominated by radio-quiet objects.

To characterize our sample, we compiled the black hole masses of these AGN from the literature.
We prioritize the black hole mass $M_{\rm BH}$ measured by reverberation mapping and then by other methods, such as the relations for the stellar velocity dispersion and the narrow-line region, as well as virial masses measured from the broad-line region.
The redshifts, R.A., and decl. coordinates are taken from NED\footnote{\url{https://ned.ipac.caltech.edu/}}.
The Galactic absorption column densities $N_{\rm H\_gal}$ toward the line of sight are taken from \citet[][]{2016A&A...594A.116H}.
The Eddington ratio ($\lambda_{\rm Edd}$) is taken from the literature \citep[e.g.][]{2009MNRAS.392.1124V,2009MNRAS.399.1553V}. 

\begin{figure*}[ht!]
\includegraphics[width=7.0in,angle=0]{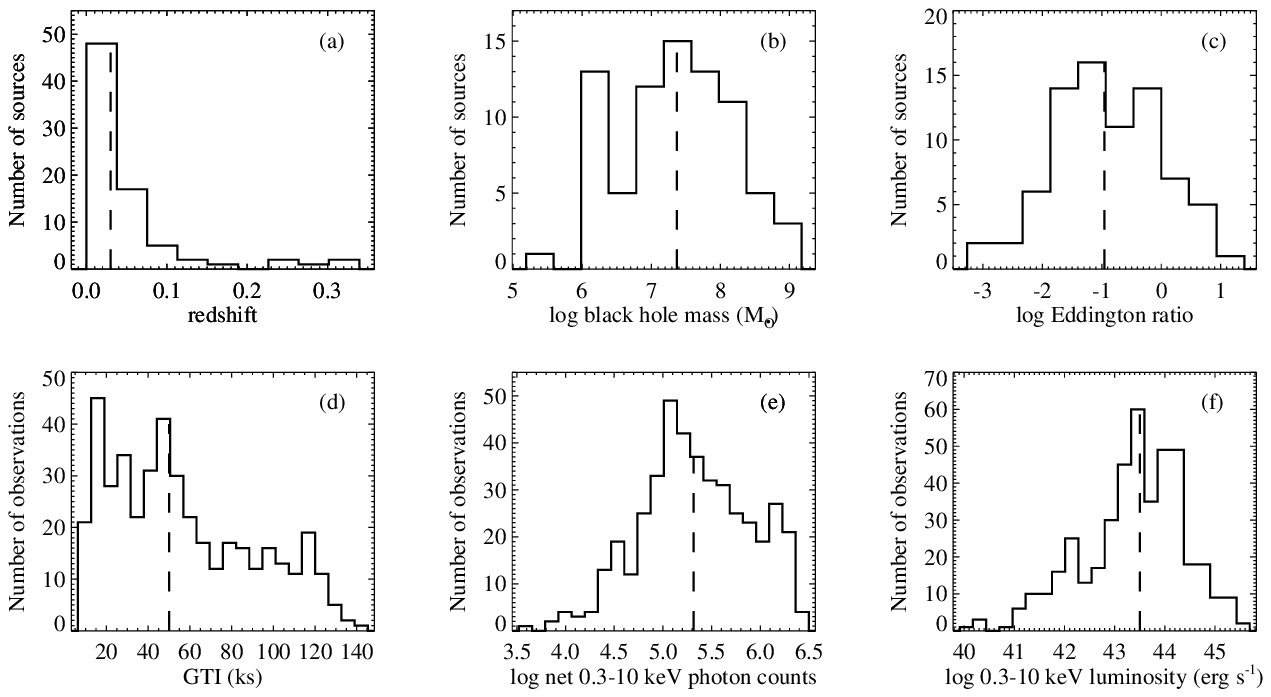}
\caption{Distributions of different properties in our sample. The dashed line denotes the median. (a) Distribution of the redshift of the sample objects. (b) Distribution of the black hole mass. (c) Distribution of the Eddington ratio. (d) Distribution of the GTI of each observation. (e) Distribution of the 0.3-10 keV net photon counts. (f) Distribution of the 0.3-10 keV luminosity.}
\label{fig:distribution}
\end{figure*}

All the parameters mentioned above are listed in Table \ref{tab:sample} for every source. The distributions of the redshift, black hole mass, Eddington ratio, good time interval (GTI), 0.3-10 keV net photon counts, and 0.3-10 keV luminosity of the selected sample are plotted in Figure \ref{fig:distribution}.
It shows that the majority (87\%) of the sample lies below redshift 0.1.
More than half of the total observations have GTI longer than 40 ks. The median values for the 0.3-10 keV net photon counts and the 0.3-10 keV luminosity are derived to be $2\times10^5$ and $3\times10^{43} \rm erg~s^{-2}$, respectively.

\begin{deluxetable*}{lccccccllc}
\tablenum{1}
\tablecaption{Table of the 78 sources in this sample, with a total of 426 observations \label{tab:sample}}
\tablewidth{0pt}
\tablehead{
\colhead{Source Name} & \colhead{Type} & \colhead{R.A.} & \colhead{Decl.} & \colhead{Redshift} & \colhead{$N_{\rm H\_gal}$} & \colhead{Number(obs)} & \colhead{${\rm log}\, M_{\rm BH} $} & \colhead{${\rm log}\, \lambda_{\rm Edd}$} & \colhead{References}\\
\colhead{} & & \colhead{(J2000.0)} & \colhead{(J2000.0)} & \colhead{} & \colhead{($10^{22}$ cm$^{-2}$)} & \colhead{} & \colhead{($M_\odot$)} & \colhead{} & \colhead{}
}
\decimalcolnumbers
\startdata
1ES 1927+654 & Sy2 & 291.8313 & 65.5651 & 0.017 & 0.064 & 3 & 7.34 & -2.23 & 1 \\
1H 0323+342 & NLS1$^{a}$ & 51.1715 & 34.1794 & 0.061 & 0.117 & 7 & 7.10 & -0.40 & 2 \\
1H 0419-577 & Sy1.5 & 66.5029 & -57.2003 & 0.104 & 0.012 & 8 & 8.11 & -0.60 & 3 \\
1H 0707-495 & NLS1 & 107.1729 & -49.5519 & 0.041 & 0.040 & 14 & 6.31 & 0.00 & 4 \\
1H 1934-063 & NLS1/Sy1.5 & 294.3875 & -6.2180 & 0.010 & 0.102 & 2 & 7.32 & -1.84 & 5 \\
3C 120 & Sy1.5$^{a}$ & 68.2962 & 5.3543 & 0.033 & 0.103 & 5 & $7.74$$^{b}$ & -0.52 & 6,7 \\
3C 390.3 & Sy1.5$^{a}$ & 280.5375 & 79.7714 & 0.056 & 0.037 & 2 & $8.46$$^{b}$ & -1.33 & 6,7 \\
Ark 120 & Sy1 & 79.0476 & -0.1498 & 0.033 & 0.100 & 6 & $8.18$$^{b}$ & -0.95 & 6,7 \\
Ark 564 & NLS1 & 340.6639 & 29.7254 & 0.025 & 0.050 & 13 & 6.06 & 0.29 & 4 \\
CTS A08.12 & Sy1.2 & 323.0090 & -33.7150 & 0.030 & 0.033 & 1 & 7.80 & -0.60 & 8,9 \\
ESO 113-G010 & NLS1/Sy1.8 & 16.3197 & -58.4373 & 0.026 & 0.020 & 1 & 6.85 & -1.25 & 10,9 \\
ESO 198-G24 & Sy1 & 39.5819 & -52.1924 & 0.045 & 0.027 & 2 & 8.28 & -1.32 & 5 \\
ESO 362-G18 & Sy1.5 & 79.8992 & -32.6576 & 0.012 & 0.013 & 2 & 7.49 & -1.21 & 5 \\
ESO 511-G030 & Sy1 & 214.8434 & -26.6447 & 0.022 & 0.043 & 1 & 7.84 & -1.58 & 11 \\
FBQS J1644+2619 & NLS1$^{a}$ & 251.1771 & 26.3203 & 0.145 & 0.050 & 1 & 7.28 & -0.79 & 12 \\
Fairall 9 & Sy1.2 & 20.9407 & -58.8058 & 0.047 & 0.029 & 6 & $8.41$$^{b}$ & -1.73 & 6,7 \\
HE 1029-1401 & Sy1 & 157.9762 & -14.2808 & 0.086 & 0.057 & 1 & 9.08 & -1.15 & 13 \\
HE 1353-1917 & Sy1 & 209.1528 & -19.5292 & 0.035 & 0.071 & 1 & 8.13 & -1.93 & 14 \\
IC 4329A & Sy1.2 & 207.3303 & -30.3094 & 0.016 & 0.041 & 3 & 8.29 & -1.38 & 11 \\
IRAS 05078+1626 & Sy1.5 & 77.6896 & 16.4988 & 0.018 & 0.189 & 1 & 7.20 & -1.40 & 5 \\
IRAS 13224-3809 & NLS1 & 201.3307 & -38.4146 & 0.066 & 0.048 & 19 & 6.82 & 0.81 & 4 \\
IRAS 13349+2438 & NLS1 & 204.3280 & 24.3843 & 0.108 & 0.011 & 4 & 8.62 & -1.64 & 15 \\
IRAS 17020+4544 & NLS1 & 255.8766 & 45.6798 & 0.060 & 0.025 & 4 & 6.77 & -0.17 & 16,17 \\
IRAS 18325-5926 & Sy2 & 279.2429 & -59.4024 & 0.020 & 0.052 & 3 & 6.36 & 0.14 & 18 \\
IRAS F12397+3333 & NLS1 & 190.5441 & 33.2841 & 0.044 & 0.014 & 2 & $6.88$$^{b}$ & -1.08 & 19,12 \\
I Zw 1 & NLS1 & 13.3956 & 12.6934 & 0.059 & 0.046 & 10 & 7.20 & 0.11 & 4 \\
MCG-02-14-009 & Sy1 & 79.0882 & -10.5615 & 0.028 & 0.101 & 1 & 7.07 & -0.92 & 20 \\
MCG-5-23-16 & Sy1.9 & 146.9173 & -30.9487 & 0.008 & 0.078 & 5 & 7.85 & -1.14 & 21 \\
MCG-6-30-15 & NLS1/Sy1.5 & 203.9738 & -34.2955 & 0.008 & 0.036 & 8 & $6.31$$^{b}$ & -0.35 & 22,23 \\
MR 2251-178 & Sy1.5 & 343.5242 & -17.5819 & 0.064 & 0.026 & 8 & 8.28 & -0.80 & 24 \\
MS 22549-3712 & NLS1 & 344.4127 & -36.9350 & 0.039 & 0.010 & 2 & 6.58 & -0.43 & 25 \\
Mrk 1040 & Sy1 & 37.0603 & 31.3117 & 0.017 & 0.067 & 3 & 6.36 & -0.73 & 4 \\
Mrk 1044 & NLS1 & 37.5230 & -8.9981 & 0.016 & 0.029 & 5 & $6.15$$^{b}$ & -0.05 & 26,3 \\
Mrk 110 & NLS1 & 141.3036 & 52.2863 & 0.016 & 0.013 & 1 & $7.40$$^{b}$ & -0.36 & 6,7 \\
\enddata
\tablecomments{The columns are: (1) the name of the source; (2) classification of the source, $^{a}$ marks the radio-loud sources, while the rest are radio-quiet objects; (3) R.A. coordinates in units of degree; (4) decl. coordinates in units of degree; (5) redshift of the source derived from the NASA/IPAC Extragalactic Database; (6) the Galactic absorption column density $N_{\rm H\_gal}$ toward the line of sight of each of the source in units of $10^{22}$  cm$^{-2}$; (7) number of observations used for each source; (8) the mass of the central black hole in units of $M_\odot$; (9) the Eddington ratio; (10) references for $M_{\rm BH}$ and $\lambda_{\rm Edd}$ values: (1) \citet{2011ApJ...726L..21T}; (2) \citet{2014ApJ...795...58L}; (3) \citet{2010ApJS..187...64G}; (4) \citet{2003MNRAS.343..164B}; (5) \citet{2007ApJ...660.1072W}; (6) \citet{2004ApJ...613..682P}; (7) \citet{2009MNRAS.392.1124V}; (8) \citet{2014AA...561A.140B}; (9) \citet{2018MNRAS.480.1522L}; (10) \citet{2013ApJ...764L...9C}; (11) \citet{2010MNRAS.402.1081V}; (12) \citet{2019ApJS..243...21L}; (13) \citet{2002ApJ...579..530W}; (14) \citet{2019AA...627A..53H}; (15) \citet{2004AA...428...39C}; (16) \citet{2001AA...377...52W}; (17) \citet{2018ApJ...867L..11L};  (18) \citet{2016AA...592A..98I}; (19) \citet{2014ApJ...782...45D}; (20) \citet{2018MNRAS.480...96W}; (21) \citet{2005ApJ...618L..83Z}; (22) \citet{2016ApJ...830..136B}; (23) \citet{2009MNRAS.399.1553V}; (24) \citet{2009AJ....137.3388W}; (25) \citet{2006ApJ...653..137Z}; (26) \citet{2014ApJ...793..108W}; (27) \citet{1998ApJ...505L..83L}; (28) \citet{2009ApJ...705..199B}; (29) \citet{2009ApJ...696..160R}; (30) \citet{2005ApJ...632..799P}; (31) \citet{2006ApJ...653..152D}; (32) \citet{2018ApSS.363..228V}; (33) \citet{2014ApJ...796....8B}; (34) \citet{2014MNRAS.445.3073P}; (35) \citet{2012MNRAS.420.1825J}; (36) \citet{2006ApJ...642..711L}; (37) \citet{2010ApJ...717.1243G}; (38) \citet{2009MNRAS.398L..16J}; (39) \citet{2008MNRAS.389.1360M}. 
$^{b}$ $M_{\rm BH}$ was measured through reverberation mapping, based on equation $M_{\rm BH}=f \frac{c \tau v^2}{G}$ with an adopted mean value $<f>=5.5$. The rest of $M_{\rm BH}$ were estimated from other methods.
}
\end{deluxetable*}

\begin{deluxetable*}{lccccccllc}
\tablenum{1}
\tablecaption{-continued \label{tab:source}}
\tablewidth{0pt}
\tablehead{
\colhead{Source Name} & \colhead{Type} & \colhead{R.A.} & \colhead{Decl.} & \colhead{Redshift} & \colhead{$N_{\rm H\_gal}$} & \colhead{Number(obs)} & \colhead{${\rm log}\, M_{\rm BH}$} & \colhead{${\rm log}\, \lambda_{\rm Edd}$} & \colhead{References} \\
\colhead{} & & \colhead{(J2000.0)} & \colhead{(J2000.0)} & \colhead{} & \colhead{($10^{22}$ cm$^{-2}$)} & \colhead{} & \colhead{($M_\odot$)} & \colhead{} & \colhead{}
}
\decimalcolnumbers
\startdata
Mrk 205 & Sy1 & 185.4343 & 75.3108 & 0.071 & 0.029 & 6 & 8.68 & -1.23 & 15 \\
Mrk 279 & Sy1 & 208.2644 & 69.3082 & 0.030 & 0.013 & 4 & $7.54$$^{b}$ & -0.67 & 6,7 \\
Mrk 335 & NLS1 & 1.5813 & 20.2029 & 0.026 & 0.033 & 8 & $7.15$$^{b}$ & -0.08 & 6,7 \\
Mrk 478 & NLS1 & 220.5310 & 35.4397 & 0.079 & 0.009 & 5 & 7.43 & -0.82 & 3 \\
Mrk 493 & NLS1 & 239.7899 & 35.0297 & 0.031 & 0.020 & 2 & $6.18$$^{b}$ & -0.06 & 26,3 \\
Mrk 509 & Sy1.5 & 311.0406 & -10.7235 & 0.034 & 0.039 & 15 & $8.16$$^{b}$ & -1.02 & 6,7 \\
Mrk 586 & Sy1.2 & 31.9578 & 2.7154 & 0.156 & 0.028 & 1 & 7.55 & 0.76 & 27 \\
Mrk 590 & Sy1 & 33.6398 & -0.7667 & 0.026 & 0.028 & 2 & $7.68$$^{b}$ & -1.98 & 6,7 \\
Mrk 704 & Sy1.2 & 139.6084 & 16.3053 & 0.029 & 0.027 & 2 & 7.76 & -0.97 & 5 \\
Mrk 766 & NLS1 & 184.6105 & 29.8129 & 0.013 & 0.018 & 9 & $6.25$$^{b}$ & -0.30 & 28,23 \\
Mrk 841 & Sy1.5 & 226.0050 & 10.4378 & 0.036 & 0.020 & 5 & 8.17 & -1.67 & 23 \\
NGC 1365 & Sy1.8 & 53.4015 & -36.1404 & 0.005 & 0.012 & 11 & 6.30 & -0.43 & 29,11 \\
NGC 1566 & Sy1.5 & 65.0017 & -54.9378 & 0.005 & 0.007 & 3 & 6.11 & -1.61 & 5 \\
NGC 2992 & Sy1.9 & 146.4252 & -14.3264 & 0.008 & 0.052 & 10 & 7.27 & -1.70 & 11 \\
NGC 3227 & Sy1.5 & 155.8774 & 19.8651 & 0.004 & 0.019 & 9 & $7.63$$^{b}$ & -3.01 & 6,7 \\
NGC 3516 & Sy1.5 & 166.6979 & 72.5686 & 0.009 & 0.031 & 6 & $7.63$$^{b}$ & -2.21 & 6,7 \\
NGC 3783 & Sy1.5 & 174.7573 & -37.7387 & 0.010 & 0.101 & 6 & $7.47$$^{b}$ & -1.40 & 6,7 \\
NGC 4051 & NLS1 & 180.7901 & 44.5313 & 0.002 & 0.012 & 19 & $6.28$$^{b}$ & -1.80 & 6,7 \\
NGC 4151 & Sy1.5 & 182.6357 & 39.4057 & 0.003 & 0.021 & 18 & $7.12$$^{b}$ & -1.16 & 6,7 \\
NGC 4395 & Sy1.8 & 186.4536 & 33.5469 & 0.001 & 0.043 & 4 & $5.56$$^{b}$ & -2.92 & 30 \\
NGC 4593 & Sy1 & 189.9143 & -5.3443 & 0.009 & 0.017 & 8 & $6.99$$^{b}$ & -1.43 & 31,7 \\
NGC 4748 & NLS1 & 193.0519 & -13.4147 & 0.015 & 0.036 & 1 & $6.41$$^{b}$ & -0.14 & 28,32 \\
NGC 5273 & Sy1.9 & 205.5345 & 35.6542 & 0.004 & 0.008 & 3 & $6.78$$^{b}$ & -2.42 & 33,12 \\
NGC 5506 & NLS1 & 213.3120 & -3.2076 & 0.006 & 0.042 & 9 & 7.67 & -1.56 & 11 \\
NGC 5548 & Sy1.5 & 214.4981 & 25.1368 & 0.017 & 0.015 & 19 & $7.83$$^{b}$ & -1.63 & 6,7 \\
NGC 6814 & Sy1.5 & 295.6692 & -10.3235 & 0.005 & 0.085 & 2 & $7.27$$^{b}$ & -2.43 & 28,34 \\
NGC 6860 & Sy1.5 & 302.1954 & -61.1002 & 0.015 & 0.030 & 1 & 7.91 & -2.07 & 11 \\
NGC 7213 & Sy1.5 & 332.3180 & -47.1666 & 0.006 & 0.011 & 2 & 7.37 & -2.11 & 11 \\
NGC 7314 & Sy2 & 338.9425 & -26.0505 & 0.005 & 0.015 & 5 & 6.14 & -1.28 & 11 \\
NGC 7469 & Sy1.5 & 345.8151 & 8.8740 & 0.016 & 0.045 & 11 & $7.09$$^{b}$ & -0.43 & 6,7 \\
NGC 985 & Sy1.5 & 38.6574 & -8.7876 & 0.043 & 0.035 & 5 & 8.36 & -1.69 & 23 \\
PG 1211+143 & NLS1 & 183.5736 & 14.0536 & 0.081 & 0.026 & 11 & $8.16$$^{b}$ & -0.62 & 6,7 \\
PG 1244+026 & NLS1 & 191.6469 & 2.3691 & 0.048 & 0.017 & 6 & 6.79 & 0.58 & 35 \\
PG 1448+273 & NLS1 & 222.7865 & 27.1574 & 0.065 & 0.030 & 3 & 7.26 & 0.46 & 35 \\
Pictor A & Sy1$^{a}$ & 79.9572 & -45.7788 & 0.035 & 0.036 & 2 & 7.60 & -0.96 & 36 \\
PKS 0558-504 & NLS1 & 89.9474 & -50.4479 & 0.137 & 0.033 & 15 & 8.40 & 0.23 & 37 \\
PKS J1220+0203 & Sy1.2$^{a}$ & 185.0495 & 2.0617 & 0.240 & 0.018 & 1 & 8.84 & -0.92 & 12 \\
RBS 229 & Sy1.2 & 25.0708 & -0.8342 & 0.334 & 0.027 & 1 & 9.01 & -1.10 & 12 \\
RE J1034+396 & NLS1 & 158.6608 & 39.6412 & 0.042 & 0.013 & 8 & 6.23 & 0.67 & 35 \\
RX J0136.9-3510 & NLS1 & 24.2267 & -35.1644 & 0.289 & 0.017 & 1 & 7.89 & 0.43 & 38 \\
RX J0439.6-5311 & NLS1 & 69.9112 & -53.1919 & 0.243 & 0.006 & 2 & 6.59 & 1.11 & 3 \\
SWIFT J2127.4+5654 & NLS1 & 321.9373 & 56.9444 & 0.014 & 0.729 & 5 & 7.18 & -0.40 & 39 \\
TON S180 & NLS1 & 14.3342 & -22.3823 & 0.062 & 0.013 & 4 & 6.85 & 0.80 & 3 \\
Ton 28 & Sy1 & 151.0109 & 28.9265 & 0.327 & 0.018 & 1 & 8.00 & -0.27 & 12 \\
\enddata
\end{deluxetable*}

\subsection{Data Reduction} \label{Data_processing}
In order to study the X-ray spectra and variability, we use the data from the {\it XMM-Newton} EPIC-pn camera as it provides the highest source count rate. The observation data files (ODF) of each observation are retrieved from the {\it XMM-Newton} Science Archive (XSA)\footnote{\url{http://nxsa.esac.esa.int}}. Then the data are reduced and analyzed using the {\it XMM-Newton} Science Analysis Software\texttt{(SAS v16.0.0)}\footnote{\url{https://www.cosmos.esa.int/web/XMM-Newton/sas-threads}}. 

Following the \texttt{SAS} data analysis threads, the \texttt{epproc} (for EPIC-pn) task is used to reprocess the data and generate new event lists. For observations with more than one calibrated and concatenated event lists of EPIC-pn, we select all useful segments to generate scientific products. We adopt a typical circular source extraction region of 40$^{\prime \prime}$ radius centered at the source coordinates.
The background is determined from a nearby source-free region, which is 50$^{\prime \prime}$ radius in the {\it PrimeSmallWindow} mode and 80$^{\prime \prime}$ radius in larger window modes. The effect of background flares are taken into account, because it may affect the analysis of variability. To do this, we visually check every background light curve and filter the event lists with the \texttt{tabgtigen} task to remove time intervals where background flares are severe. For the purpose of reliable timing analysis, we only use continuous GTIs of more than 10 ks long (Figure \ref{fig:distribution}(d)).

For the EPIC-pn camera, $\sim$66.9\% of the observations were operated in the {\it PrimeSmallWindow} mode, while the others were in the {\it PrimeFullWindow} mode ($\sim$14.3\%), the {\it PrimeFullWindowExtended} mode ($\sim$0.5\%) and the {\it PrimeLargeWindow} mode ($\sim$18.3\%), thus photon pileup can be a problem for bright sources. Therefore, we check the pileup effect by running the \texttt{epatplot} task and minimize it by excising the core of the point spread function (PSF) in case of need, i.e. replacing the circular region with an annular region.

Within the selected GTIs, both light curves and time-averaged X-ray spectra are extracted from the source and background regions with the \texttt{evselect} task. Only good events with PATTERN $\leq$ 4 are adopted for analysis.
The light curves are binned with 100 s per bin.
The background-subtracted source light curves are produced with the \texttt{epiclccorr} task.
The redistribution matrix and ancillary file are produced for each spectrum with the \texttt{SAS} tasks \texttt{rmfgen} and \texttt{arfgen}.
All the spectra are rebinned with the FTOOLS task \texttt{grppha}, so that each bin contains at least 25 counts. \texttt{XSPEC v12.9.1} is used to perform all the spectral fitting.

In order to measure the optical/UV flux of each source in every observation, we also reduce all the data from the {\it XMM-Newton} Optical Monitor (OM). The OM data are reprocessed with the \texttt{omichain} task.

The rms variability amplitude from X-ray light curves is calculated by adopting the frequency-resolved spectral techniques. The calculation procedure is summarized in Appendix~\ref{sec-calRms} \citep[see also][]{2003MNRAS.345.1271V, 2008MNRAS.387..279A,2008MNRAS.389.1427P,2017MNRAS.468.3663J}. We emphasize that the subtraction of the Poisson noise power can affect the accuracy of the rms significantly. We find that the theoretical Poisson noise power can significantly underestimate the true Poisson noise, which then leads to the overestimate of the intrinsic rms, and so further corrections should be applied. We suggest that the arithmetic average of the power over the power spectra density (PSD) high-frequency range, which is dominated by the Poisson noise power is the most accurate method in this work. A detailed prescription of the determination of the Poisson noise power can be found in Appendix~\ref{PoissonNoise}.

\section{The Diverse RMS and Time-averaged Spectra: A Statistic View}
\label{sec-statistics}
\subsection{Production of the Rms and Time-averaged Spectra}
A time-averaged spectrum shows the source flux in every energy bin. In comparison, an rms spectrum shows fractional flux variation in every energy bin (see \citealt{2014A&ARv..22...72U} and references therein). The amplitude of variation is also dependent on the variation frequency (i.e. timescale; \citealt{2013MNRAS.436.3173J,2017MNRAS.468.3663J}). Therefore, a joint analysis of the time-averaged spectrum and rms spectrum will allow the exploration of different spectral components with different variability properties.

We calculate the rms spectra for the entire sample of 78 Seyfert galaxies and 426 {\it XMM-Newton} observations.
In order to increase the accuracy of rms measurement, we adopt a careful selection of seven energy bins spanning 0.3-10 keV, in order to keep sufficient photon counts per bin and to produce reliable power spectra. These energy bins are selected to reach a reasonable compromise between the spectral resolution and the accuracy of rms measurement in every energy bin. In some cases the source flux is still too low to produce reliable rms in individual energy bins, then we merge adjacent bins to increase the count rates, at the expense of losing more energy resolution. On the other hand, for a subset of bright sources, the energy resolution of the rms spectra can be increased to allow a more detailed study. However, for the consistency of rms spectra production and ease of shape comparison for the entire sample, we do not optimize the energy resolution for every source. These bright sources can be investigated in more detail on a case by case basis (e.g. \citealt{2020MNRAS.492.1363P,2021MNRAS.500.4506H}).

In addition, we divide the frequency range into the low-frequency (LF; $<10^{-4}$ Hz) and high-frequency (HF; $10^{-4}$ - $10^{-3}$ Hz) bands, in order to explore the frequency (i.e. timescale) dependence of the rms spectra. To this end, observations without complete fractional LF or HF rms spectra across 0.3-10 keV (due to unconstrained rms in some energy bins) are discarded, while those with high-quality rms spectra are culled, including 330 observations of 69 sources.

For each of these sources, we extract the time-averaged spectrum in every observation. Then these 330 time-averaged spectra are fitted in 2-10 keV with a single power-law model absorbed by a Galactic column and a free intrinsic column. The absorption is modeled by the {\tt (z)TBabs} model (\citealt{2000ApJ...542..914W}), and the Galactic column is fixed at the value along the source's line of sight (\citealt{2016A&A...594A.116H}). Then we produce the unfolded time-averaged spectra for the following analysis.

\begin{figure*}[ht!]
\centering
\includegraphics[width=7.0in]{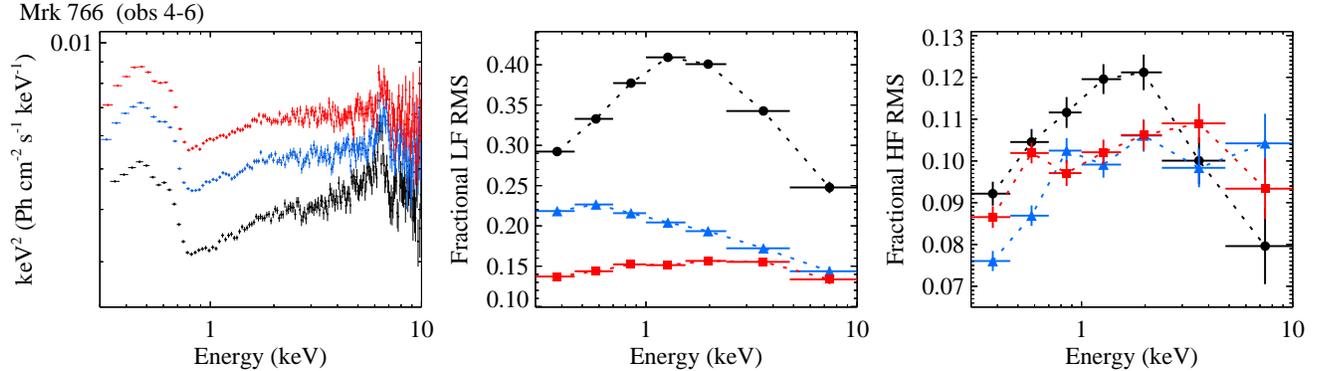}
\caption{Typical example of the results of the connection of the rms spectra with the time-averaged spectra for the case of Mrk 766. Similar plots for other sources in Table \ref{tab:sample} can be found in the appendix. The left panel shows the unfolded time-averaged spectra of three separate observations. The middle panel shows the fractional rms of each individual observation in the low frequency (LF; $<10^{-4}$ Hz). The right panel shows the fractional rms in the high frequency (HF; $10^{-4}$ - $10^{-3}$ Hz). }
\label{fig:rmsspec}
\end{figure*}

\subsection{A Typical Example: Mrk 766}
Firstly, we show a typical example from the above analysis, which is Mrk 766 shown in Figure \ref{fig:rmsspec}.
The unfolded time-averaged spectra from three separate observations are shown in the left panel. These observations were conducted between 2005 May 25 and 2005 May 29. These spectra mainly differ in their overall fluxes, while their shapes are all similar, including a power-law-shape continuum above 2 keV, a significant Fe K$\alpha$ emission line in 6-7 keV and a strong soft X-ray excess below 0.8 keV.

The fractional rms spectra of each individual observation in LF and HF are shown in the middle and right panel, respectively.
The LF rms spectra show various spectral shapes in different observations.
The concave-down shape of the LF rms spectrum in black indicates that the soft excess and the hard X-ray component are much less variable than the middle energy band of 1-2 keV. This can be qualitatively explained by the classic disk-reflection model \citep[e.g.][]{2004MNRAS.349.1435M,2005MNRAS.358..211R,2013MNRAS.429.2917F}, where the soft excess and hard X-ray is dominated by the relatively constant reflection component, while the more variable primary power law dominates the energy band in the middle. 
\citet{2020MNRAS.492.1363P} presented simple \texttt{XSPEC} models for fitting the excess variance spectra and found that the rms spectra of IRAS 13224-3809 can be well described by a variable power law damped with a constant soft excess and a less-variable relativistic reflection component, enhanced with a ultrafast outflow (UFO) in that case. Alternatively, this can also be qualitatively explained by the ionized absorption model, where the variability of the ionized absorption introduces a peak of the rms in the energy band round 1 keV \citep{2021MNRAS.508.1798P}.

However, Figure \ref{fig:rmsspec} also shows that the situation is much more complicated. We see that as the flux of the time-averaged spectra increases, the overall LF rms decreases. This is can be understood as if there is a constant component contributing more flux to the higher-flux time-averaged spectra across the entire 0.3-10 keV. However, then we also observe a significant change in the shape of the LF rms spectra as the flux increases, which is difficult to explain in the reflection scenario. Another possibility is that the low-flux spectrum is more absorbed and the absorber can introduce short-term variability \citep{2014MNRAS.442.2456G}, but this also requires more detailed quantitative analysis of all the observations of Mrk 766, which is beyond the scope of this work. On the other hand, the HF fractional rms spectra are roughly similar, which mainly follow the concave-down shape.

A linear relation between the rms variability and flux was observed in X-ray binaries (XRBs) and AGN \citep{2001MNRAS.323L..26U,2004ApJ...612L..21G}, suggesting that variations on different timescales must be coupled together \citep{2005MNRAS.359..345U}.
For each observation of Mrk 766 in our sample, we also find a linear relation between the average absolute rms amplitude and the flux on short timescales. This is consistent with previous results by using 2001 {\it XMM-Newton} data \citep{2003MNRAS.345.1271V} and 2005 {\it XMM-Newton} data \citep{2007ApJ...656..116M}. We also explore this relation on long timescales. Although three observations shown in Figure \ref{fig:rmsspec} display an anticorrelation between the LF rms and flux, the total nine observations of this source exhibit a positive linear rms-flux correlation. A more complete analysis about this rms-flux relation of our sample will be presented in a following paper (Hu et al. 2022, in preparation).

\begin{figure*}[ht!]
\includegraphics[width=7.0in]{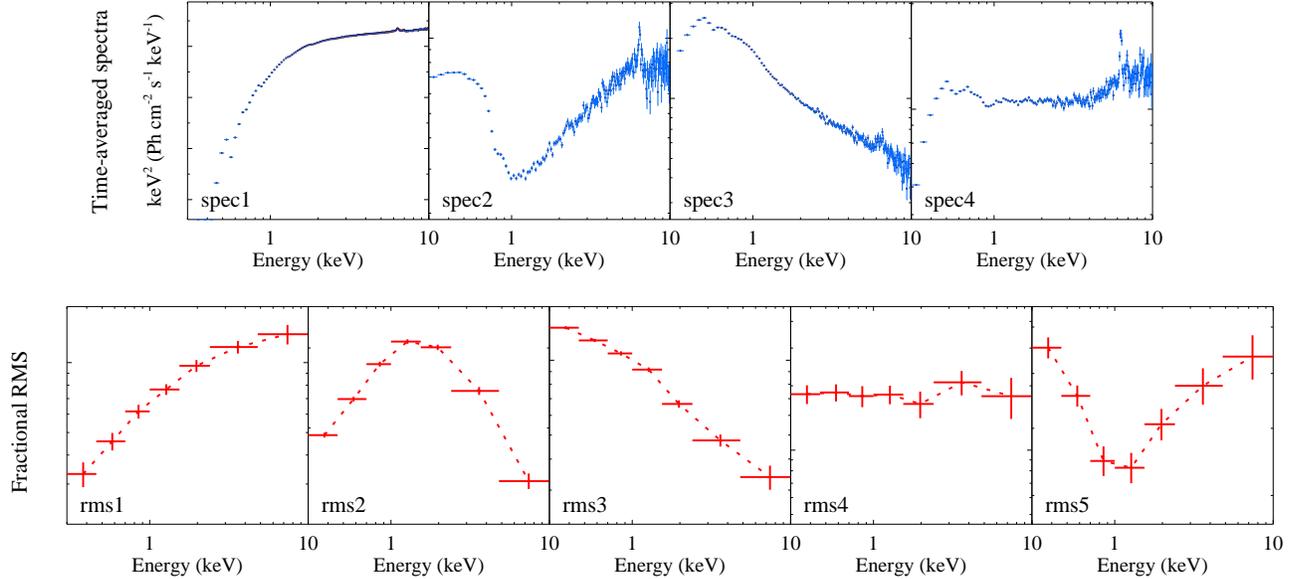}
\caption{Type of the spectral shape of the time-averaged spectra and the fractional rms spectra. The time-averaged spectra,  plotted in $E^2N(E)$, can be classified as: (spec1) increasing with energy; (spec2) concave-up; (spec3) decreasing with energy; (spec4) flat. The LF and HF rms spectral shapes can be classified as: (rms1) increasing with energy; (rms2) concave-down; (rms3) decreasing with energy; (rms4) flat; (rms5) concave-up.}
\label{fig:spec_type}
\end{figure*}

The example of Mrk 766 clearly shows that there is a complex diversity in the shape of the rms spectrum, and its link to the time-averaged spectrum is also complicated. More results for other sources are shown in Figure \ref{src_rmsspec} in the appendix. In order to obtain a global understanding of the diversity and complexity of the rms and time-averaged spectra, we perform a statistical analysis of these spectra using the entire sample of 69 Seyferts and 330 observations.

\subsection{Statistical Results}
\begin{figure*}
\includegraphics[width=6.8in]{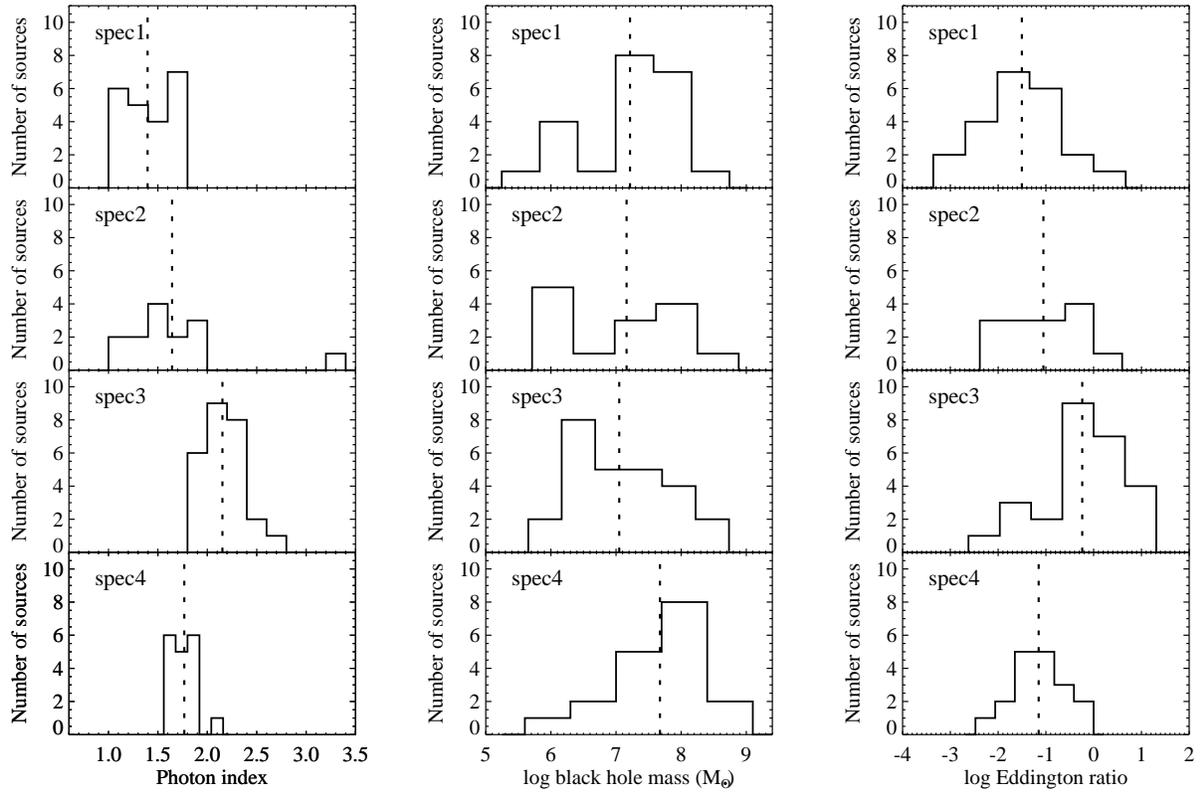}
\caption{Distribution of the photon index, black hole mass, and Eddington ratio of each source in different time-averaged spectral subgroups. The vertical dashed lines represent the average values.}
\label{fig:distribution_group}
\end{figure*}

\begin{figure*}
\includegraphics[width=6.8in]{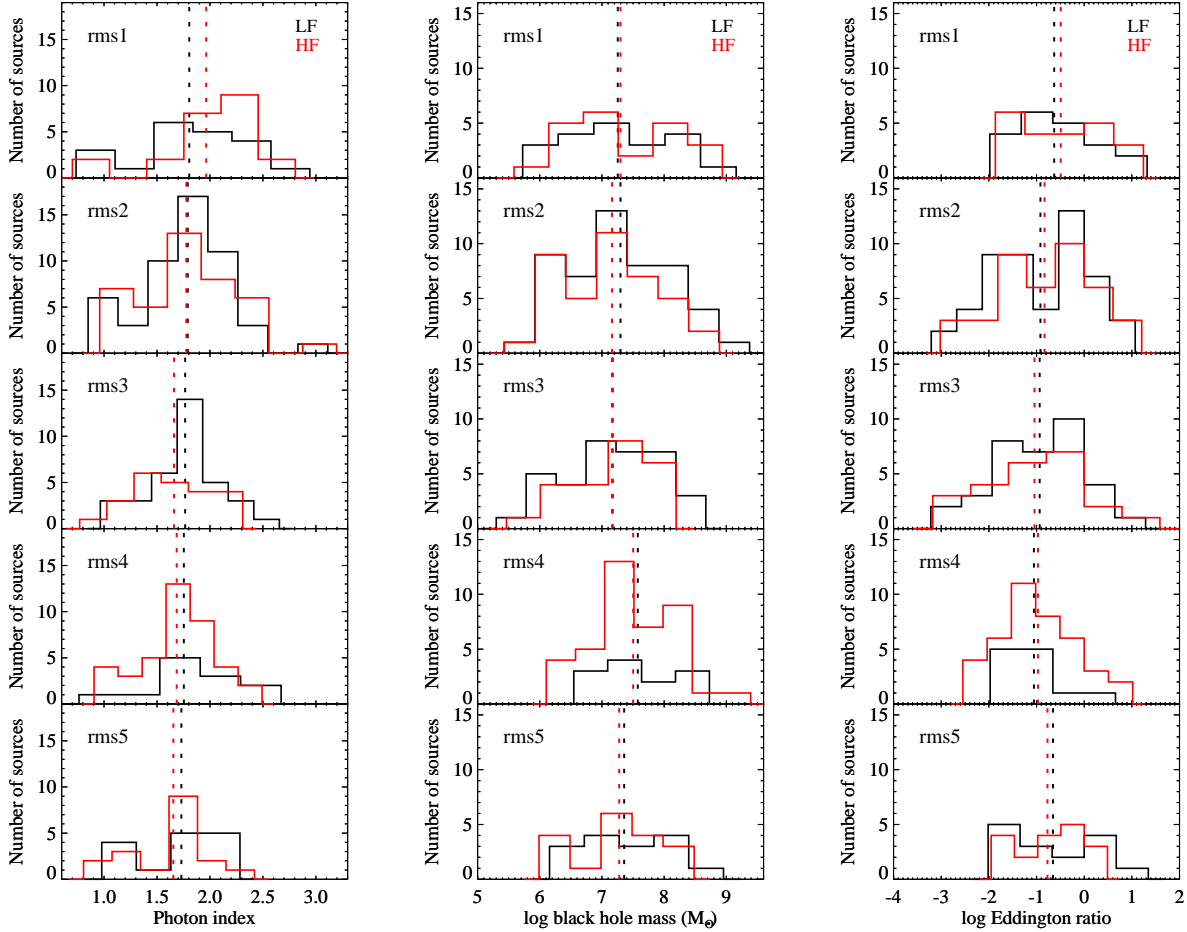}
\caption{Distribution of the photon index, black hole mass, and Eddington ratio of each source in different rms spectral subgroups. The vertical dashed lines represent the average values. The LF rms subgroups are shown in black lines, and the HF rms subgroups are shown in red lines.}
\label{fig:distribution_group_rms}
\end{figure*}

Firstly, we visually inspect all the rms and time-averaged spectra. Then based on the observed spectral shapes, we classify the spectra into several subsamples, as shown in Figure \ref{fig:spec_type}. For the time-averaged spectra plotted in $E^2N(E)$, which is unfolded through a power-law model, the spectral shapes are classified as: (spec1) increasing with energy; (spec2) concave-up; (spec3) decreasing with energy; (spec4) flat. In addition, the fractional LF and HF rms spectral shapes are classified as: (rms1) increasing with energy; (rms2) concave-down; (rms3) decreasing with energy; (rms4) flat; (rms5) concave-up. Then we calculate the mean value of some key parameters within each subsample, including the hard X-ray photon index ($\Gamma_{\rm 2-10\,keV}$), black hole mass ($M_{\rm BH}$), and Eddington ratio ($\lambda_{\rm Edd}$). The results are presented in Table \ref{tab:spec_info}. The distributions of the photon index, black hole mass $M_{\rm BH}$, and Eddington ratio $\lambda_{\rm Edd}$ for every subsample are plotted in Figs. \ref{fig:distribution_group} and \ref{fig:distribution_group_rms}. Note that one source may appear in multiple subsamples, if it shows more than one type of spectral shape in different observations.

For the time-averaged spectra, we find that the spec3 group has the steepest mean $\Gamma_{\rm 2-10\,keV}$ of 2.15. Its mean $\lambda_{\rm Edd}$ of 0.575 is also the largest among the four groups. This is because many Seyferts in the spec3 group are super-Eddington NLS1s (see Figure \ref{fig:distribution_group}), such as PG 1244+026 \citep{2013MNRAS.436.3173J}, RX J0439.6-5311 \citep{2017MNRAS.471..706J}, and RX J0136.9-3510 \citep{2009MNRAS.398L..16J}. This group also has the smallest mean $M_{\rm BH}$, as expected for these NLS1s with $M_{\rm BH}\lesssim10^{7}M_{\sun}$. On the contrary, the spec1 group has the smallest $\Gamma_{\rm 2-10\,keV}$ of 1.40, and its mean $\lambda_{\rm Edd}$ is only 0.032. This is consistent with the known correlation found between the hard X-ray photon index and Eddington ratio in various AGN samples \citep[e.g.][]{1999ApJ...526L...5L,2004ApJ...607L.107W,2008ApJ...682...81S,2011ApJ...727...31A,2011ApJ...733...60T,2012MNRAS.420.1825J,2018MNRAS.477..210Q, 2019MNRAS.487.3884C}. It is also interesting to note that the spec4 group has the largest mean $M_{\rm BH}$ of $4.8\times10^{7}M_{\sun}$ and small mean $\lambda_{\rm Edd}$ of 0.071.

Among the five HF rms groups, the HF rms1 group shows the largest mean $\Gamma_{\rm 2-10\,keV}$ of 1.96 in their corresponding time-averaged spectra and the largest mean $\lambda_{\rm Edd}$ of 0.324. This means that sources with higher Eddington ratios tend to show stronger HF variability in the hard X-ray band. In comparison, the lowest mean $\lambda_{\rm Edd}$ of 0.091 is found in the HF rms3 group. The smallest mean $\Gamma_{\rm 2-10\,keV}$ of 1.66 is found in the HF rms3 and rms5 group, both of which show increasing fractional rms toward lower energy below 1 keV. 

Among the five LF rms groups, we find that the LF rms1 group shows the largest mean $\Gamma_{\rm 2-10\,keV}$ of 1.80 and the largest mean $\lambda_{\rm Edd}$ of 0.234, similar to the HF rms1 group. The lowest mean $\lambda_{\rm Edd}$ of 0.087 is found in the LF rms4 group. The low $\lambda_{\rm Edd}$ of the LF and HF rms4 groups indicates that sources with low Eddington ratios tend to show similar fractions of variability across the 0.3-10 keV in both the LF and HF bands.

The above results suggest the likely presence of systematic links between the rms and time-averaged spectra, which are described in the next section.

\section{Connections between the rms and time-averaged spectra}
\label{sec-connection}
The AGN's X-ray spectral fitting is often degenerated to different models, because a range of physical processes can affect the spectral shape. For example, multiple components can contribute to the X-ray spectrum at the same time, such as the disk blackbody emission, warm/hot corona Comptonization, neutral/ionized reflection, emission from environmental gas. Also, the spectral shape can be modified by absorbers along the line of sight with different velocities, ionization states, and covering factors. In addition, the strong gravity around the SMBH and the disk wind materials with different velocities can provide smearing mechanisms for the emission/absorption lines. Therefore, a joint analysis of the rms spectra is especially important because it can provide additional constraints on the X-ray production mechanism \citep[e.g.][]{2011MNRAS.417..250M, 2013MNRAS.436.3173J, 2014MNRAS.437..721P, 2015MNRAS.447...72P, 2021MNRAS.500.2475J, 2021MNRAS.508.1798P}. The large sample of AGN with high-quality data presented in this work enables us, for the first time, to explore the possible systematic links between the rms spectra and time-averaged spectra.

\begin{figure*}[ht!]
\includegraphics[width=7.1in]{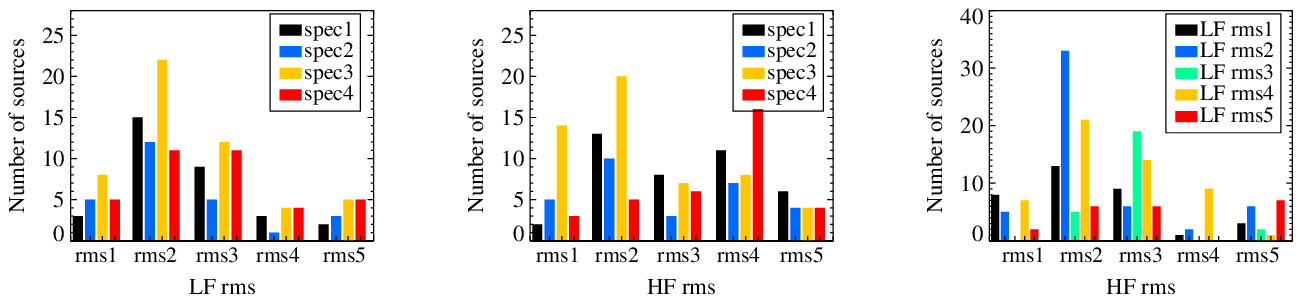}
\caption{Connection of the rms spectrum and the time-averaged spectrum. Left panel: distribution of the time-averaged spectral classification in different LF rms subsamples. Middle panel: distribution of the time-averaged spectral classification in different HF rms subsamples. Right panel: distribution of the LF rms classification in different HF rms subsamples.}
\label{fig:connection}
\end{figure*}

As shown in Table \ref{tab:spec_rms}, for each of the time-averaged spectral groups from spec1 to spec4, we calculate the proportions of different rms spectral groups. If several observations of one source show the same type of the time-averaged spectra and the same type of the rms spectra, then they are counted only once. If a source shows one type of time-averaged spectrum but multiple types of rms spectra, then it is counted in the calculation of proportion for those types of rms spectra. To visualize the links, the distributions of different types of the time-averaged spectra within every rms spectral group are plotted in Figure \ref{fig:connection}. These table and figure show a range of systematic connections between the time-averaged spectra and rms spectra, as below.

(a) The most common rms spectral shape is rms2, where the fractional rms peaks in the energy band of 1-2 keV and decreases toward both soft and hard X-rays. Rms2 dominates the spectral groups from spec1 to spec3. This type of rms spectral shape can be modeled with the ionized reflection model, where the middle energy band contains the highest fraction of the primary continuum emission, which is most variable \citep[e.g.][]{2015MNRAS.449..234K,2021MNRAS.508.1798P}. The only exception is the spec4 group, where the spectral shape is relatively flat. This group prefers a relatively flat HF rms spectrum, i.e .the HF rms4, although in the LF band this group is still dominated by the LF rms2.

(b) The rarest rms spectral shape is rms5, where both the soft and hard X-rays are more variable than the energy band of 1-2 keV. 
This energy band is sensitive to the emission lines which can be less variable than the continuum \citep[e.g.][]{2021MNRAS.506.5190L}. The dip can be due to the photoionized gas emission on large scales (e.g. narrow-line region, galactic scales), and it is found that a strong damping feature around 1 keV in proper ionization can exist in the rms spectra \citep{2021MNRAS.508.1798P}, so this might explain the few cases where such an rms spectral shape is observed.

\begin{deluxetable*}{cccccccc}
\tablenum{2}
\tablecaption{Properties of the Selected Samples 
\label{tab:spec_info}}
\tablewidth{0pt}
\tablehead{
Spectral shape & $N_{\rm src\_all}$ & $<$Redshift$>$ & $<\Gamma_{\rm 2-10\,keV}>$ & $<{\rm F_{0.3-10\,keV}}>$ & ${<\rm L_{0.3-10\,keV}}>$ & $<{\rm log}\, M_{\rm BH}>$ & $<{\rm log}\, \lambda_{\rm Edd}>$ \\
 & &  &  & ($10^{-11}\ \rm erg\,s^{-1}\,cm^{-2}$) & ($10^{43}\ \rm erg\, s^{-1}$) & ($M_{\odot}$) & 
 }
\decimalcolnumbers
\startdata
spec1 & 22 & 0.017 & 1.40 & 4.8 & 2.0 & 7.21 & -1.50 \\
spec2 & 14 & 0.029 & 1.64 & 2.6 & 3.9 & 7.16 & -1.05 \\
spec3 & 26 & 0.073 & 2.15 & 2.3 & 25.3 & 7.05 & -0.24 \\
spec4 & 18 & 0.034 & 1.77 & 4.6 & 14.3 & 7.68 & -1.15 \\
\hline
HF rms1 & 22 & 0.056 & 1.96 & 3.2 & 23.6 & 7.30 & -0.49 \\
HF rms2 & 40 & 0.047 & 1.78 & 3.3 & 14.8 & 7.16 & -0.83 \\
HF rms3 & 23 & 0.021 & 1.66 & 3.7 & 4.6 & 7.16 & -1.04 \\
HF rms4 & 40 & 0.035 & 1.69 & 3.9 & 13.6 & 7.50 & -0.97 \\
HF rms5 & 18 & 0.033 & 1.66 & 3.9 & 14.0 & 7.28 & -0.77 \\
\hline
LF rms1 & 20 & 0.052 & 1.80 & 3.6 & 14.7 & 7.25 & -0.63 \\
LF rms2 & 51 & 0.044 & 1.79 & 3.8 & 15.0 & 7.30 & -0.92 \\
LF rms3 & 35 & 0.029 & 1.77 & 3.5 & 5.8 & 7.17 & -0.93 \\
LF rms4 & 12 & 0.038 & 1.75 & 4.0 & 12.7 & 7.58 & -1.06 \\
LF rms5 & 15 & 0.043 & 1.73 & 3.1 & 17.4 & 7.36 & -0.66 \\
\enddata
\tablecomments{The columns are: (1) different time-averaged spectral shape and rms spectral shape (spec1$\sim$spec4, rms1$\sim$rms5); (2) total number of source, several observations of one source showing the same type of the spectral shape are classified as one source; (3) average redshift; (4) average photon index; (5) average 0.3-10 keV flux in units of $10^{-11}\ \rm erg\ s^{-1}\ cm^{-2}$; (6) average  0.3-10 keV luminosity in units of $10^{43}\ \rm erg\ s^{-1}$; (7) average logarithmic black hole mass in units of $M_{\odot}$; (8) average logarithmic Eddington ratio.}
\end{deluxetable*}

\begin{deluxetable*}{ccccccccccccc}
\tablenum{3}
\tablecaption{Proportion of Different Subsamples
\label{tab:spec_rms}}
\tablewidth{0pt}
\tablehead{
Time-averaged & \multicolumn6c{Fractional HF rms Spectra}&\multicolumn6c{Fractional LF rms Spectra} \\
\cmidrule(r){2-7} \cmidrule(r){8-13}
\noalign{\smallskip}
spectra & rms1 & rms2 & rms3 & rms4 & rms5 & $N_{\rm src}$ & rms1 & rms2 & rms3 & rms4 & rms5 & $N_{\rm src}$
 }
\decimalcolnumbers
\startdata
spec1 & 5.0\% & 32.5\% & 20.0\% & 27.5\% & 15.0\% & 40 & 9.4\% & 46.9\% & 28.1\% & 9.4\% & 6.2\% & 32 \\
spec2 & 17.2\% & 34.5\% & 10.3\% & 24.1\% & 13.8\% & 29 & 19.2\% & 46.2\% & 19.2\% & 3.8\% & 11.5\% & 26 \\
spec3 & 26.4\% & 37.7\% & 13.2\% & 15.1\% & 7.5\% & 53 & 15.7\% & 43.1\% & 23.5\% & 7.8\% & 9.8\% & 51 \\
spec4 & 8.8\% & 14.7\% & 17.6\% & 47.1\% & 11.8\% & 34 & 13.9\% & 30.6\% & 30.6\% & 11.1\% & 13.9\% & 36 \\
\hline
spec all & 15.4\% & 28.0\% & 16.1\% & 28.0\% & 12.6\% & 143 & 15.0\% & 38.3\% & 26.3\% & 9.0\% & 11.3\% & 133 \\
\enddata
\tablecomments{The definition of each spectral shape are described in Figure \ref{fig:spec_type}. The columns are: (1) different time-averaged spectral shape (spec1$\sim$spec4); (2)-(7) the proportion of each fractional high-frequency(HF) rms spectral shape (rms1$\sim$rms5) in total number of source, several observations of one source showing the same type of the time-averaged spectra and the same type of the rms spectra are classified as one source; (8)-(13) same as column 2-7, but for fractional low-frequency(LF) rms spectra.}
\end{deluxetable*}

(c) The spec1 group appears most severely absorbed in the soft X-rays. Interestingly, we find that the second most preferred rms spectral shape of this group is rms4 in the HF band, which is relatively flat. This means that in 27.5\% sources of this spectral group, the absorption does not increase or suppress the HF variability in the soft X-ray band. In comparison, the second most preferred rms in the LF band is rms3, which shows an enhanced fractional variability toward soft X-rays. This may suggest that either the absorption introduces extra LF variability \citep[e.g.][]{2021MNRAS.508.1798P}, or that the spectral components with less LF variability are more strongly absorbed.

(d) The shape of the spec2 group is mainly dominated by rms2. The shapes of the time-averaged spectra and rms spectra are qualitatively consistent with both the ionized reflection model and warm absorption model. Some sources in this group has been studied in great details, such as 1H 0707-495 \citep[e.g.][]{2021MNRAS.508.1798P} and IRAS 13224-2809 \citep[e.g.][]{2020MNRAS.492.1363P,2020NatAs...4..597A}.

(e) The shape of the spec3 group is typical for some ``X-ray simple'' NLS1s \citep{2006MNRAS.368..479G}. We note that the second most preferred rms spectral shape of this group is rms1 in the HF band and rms3 in the LF band. These results are consistent with previous studies on some super-Eddington NLS1s such as PG 1244+026 \citep{2013MNRAS.436.3173J} and RX J0439.6-5311 \citep{2017MNRAS.468.3663J}, where the soft excess is dominated by a separate warm Comptonization component, which is more variable than the hard X-rays in the LF band but less variable in the HF band. In the same scenario, the rms2 shape can be understood if in those cases the soft X-ray excess component is not variable, and the hard X-rays contain a less-variable neutral reflection component.

(f) The flattest time-averaged spectra of the spec4 group more difficult to explain. Different combinations of physical processes may be responsible for this spectral shape. We find that in the HF band this spectral group prefers the flat rms spectral shape of rms4, while in the LF band it prefers both rms2 and rms3. Therefore, rms spectra can be used to distinguish different physical models producing similar flat time-averaged spectra.

(g) By comparing the HF and LF rms spectral groups, we find that in most cases the LF and HF rms spectra have similar shapes, which is clearly revealed in Figure~\ref{fig:connection} right panel. Individual observations can show different combinations of the LF and HF rms spectral shapes. This indicates that the variability properties of individual sources can be very different and complicated.

For some of the systematic connections mentioned above, we have discussed briefly possible physical scenarios which were proposed by previous studies of individual sources.
However, this does not mean that these scenarios are exclusive. Indeed, there are different physical scenarios to explain the same combination of spectral shape and variability. For example, both disk-reflection model and wind absorption model are able to explain the spectral-timing properties of 1H 0707-495 \citep[e.g.][]{2015MNRAS.449..234K,2016MNRAS.460.1716D,2016MNRAS.461.3954H,2021MNRAS.508.1798P}. The verification of different X-ray models requires detailed spectral-timing analysis of individual sources, which is beyond the scope of this work.

\begin{figure*}[ht!]
\includegraphics[width=1.75in]{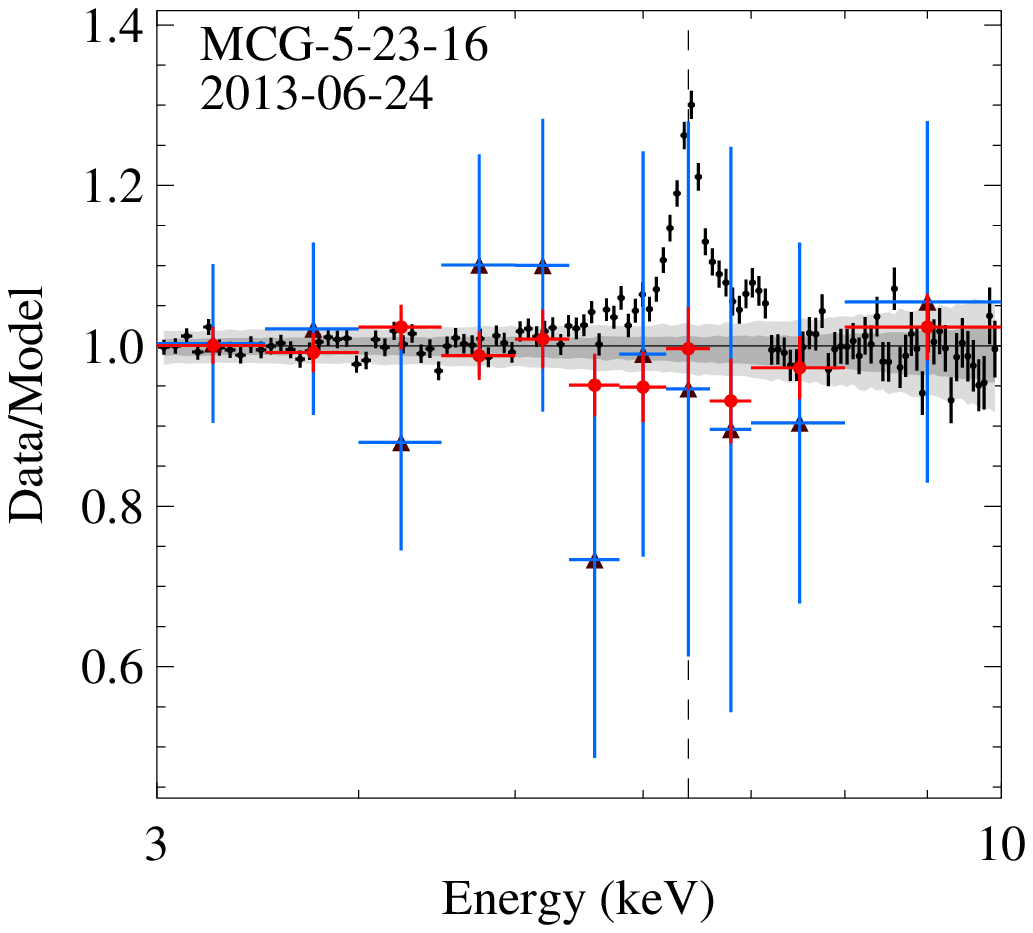} 
\includegraphics[width=1.75in]{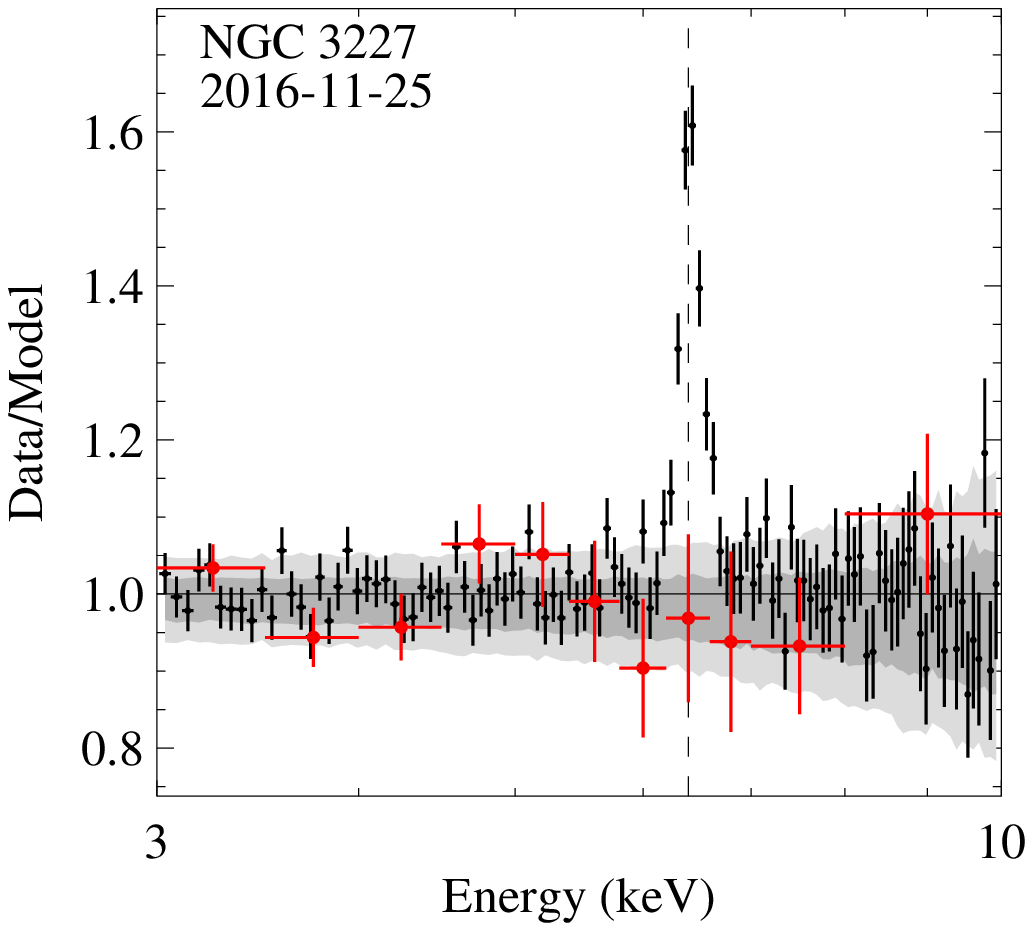} 
\includegraphics[width=1.75in]{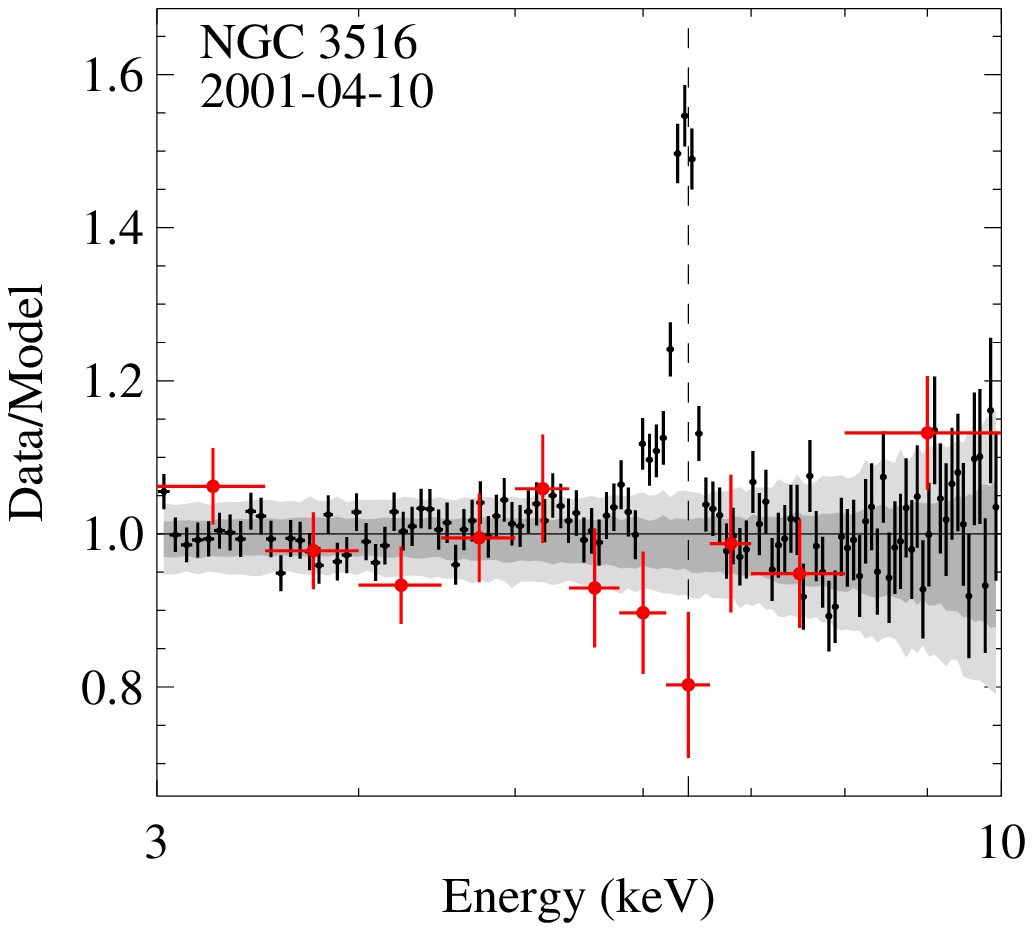} 
\includegraphics[width=1.75in]{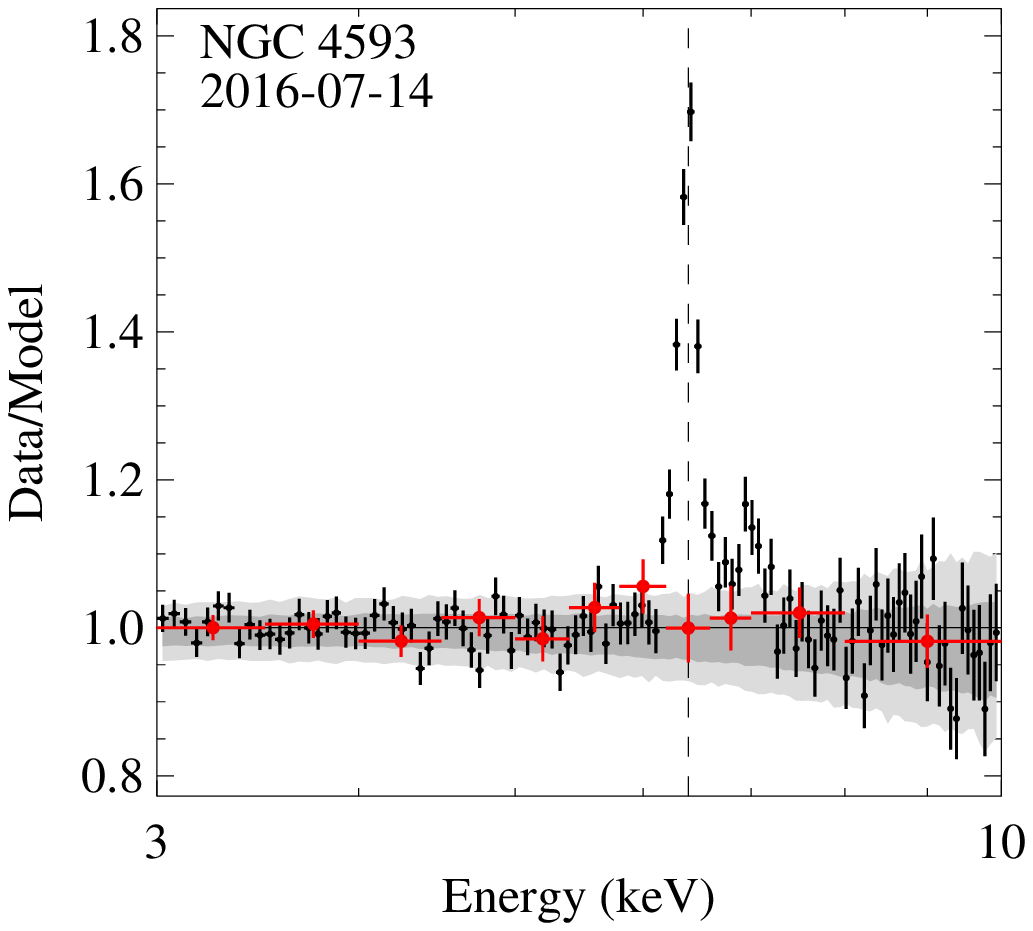} 
\includegraphics[width=1.75in]{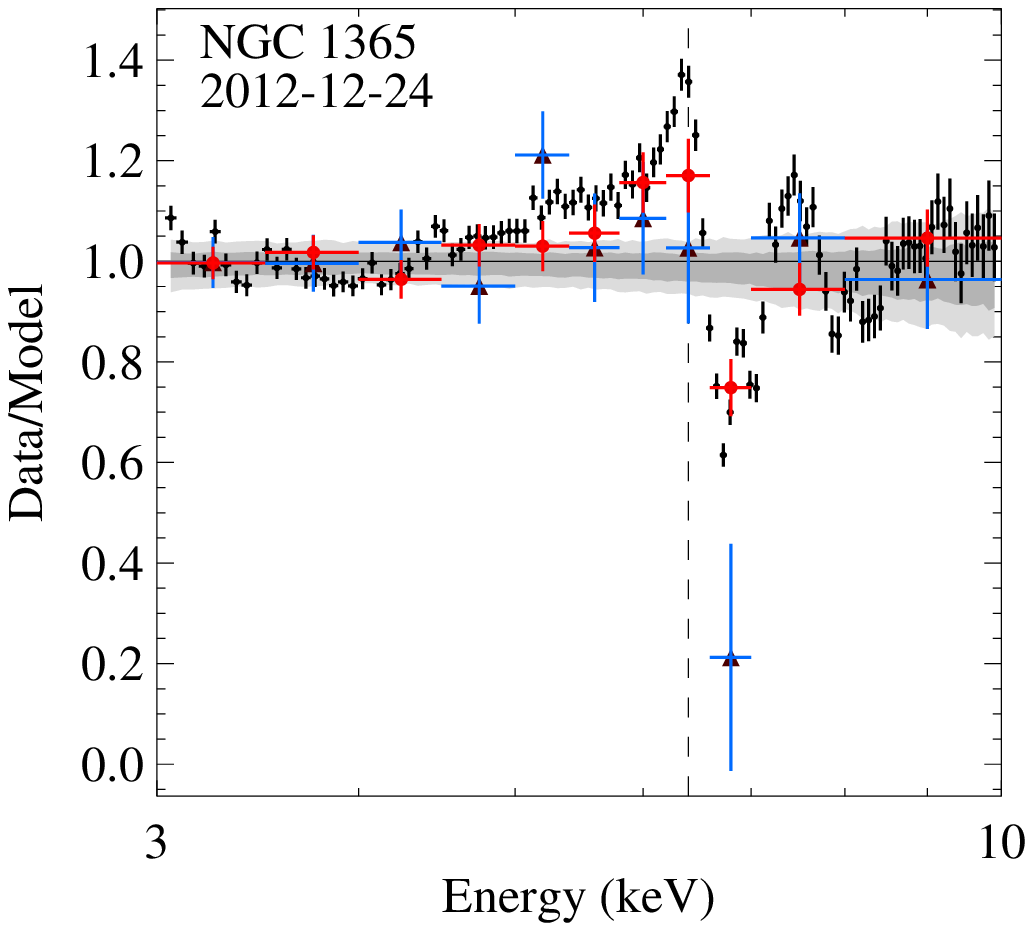} 
\includegraphics[width=1.75in]{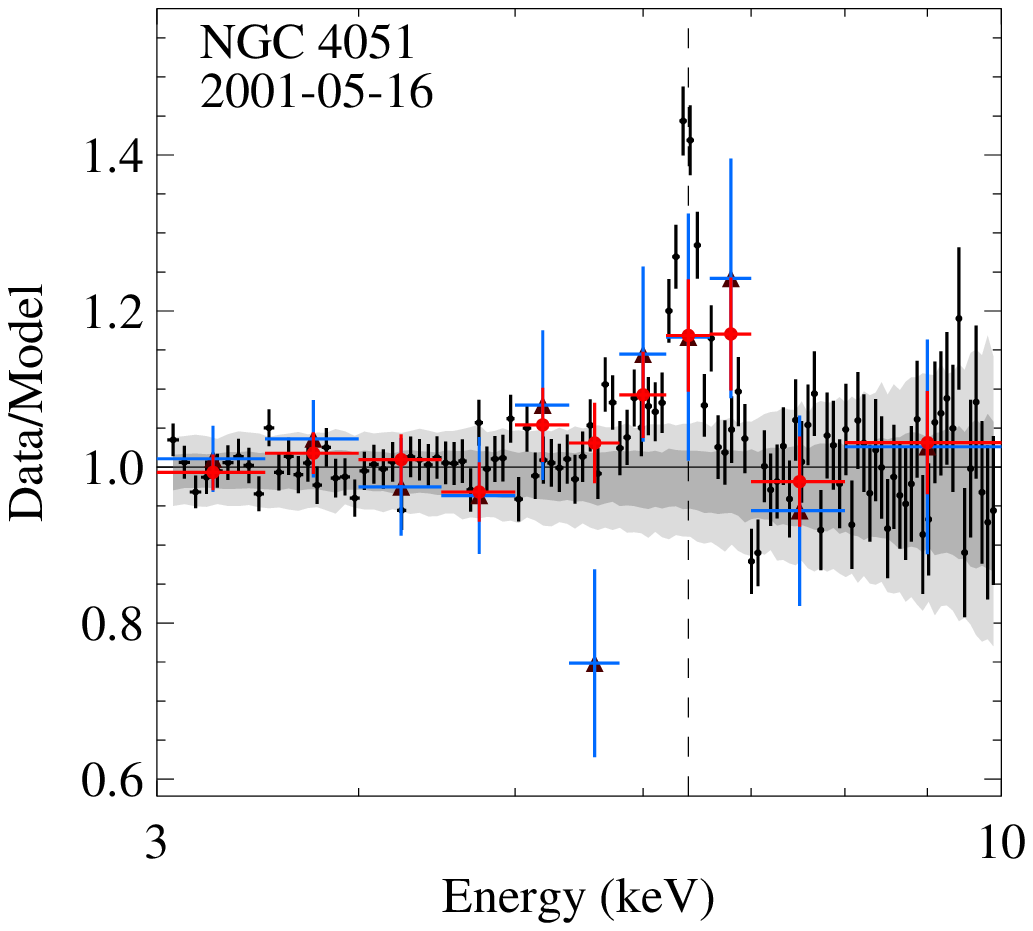} 
\includegraphics[width=1.75in]{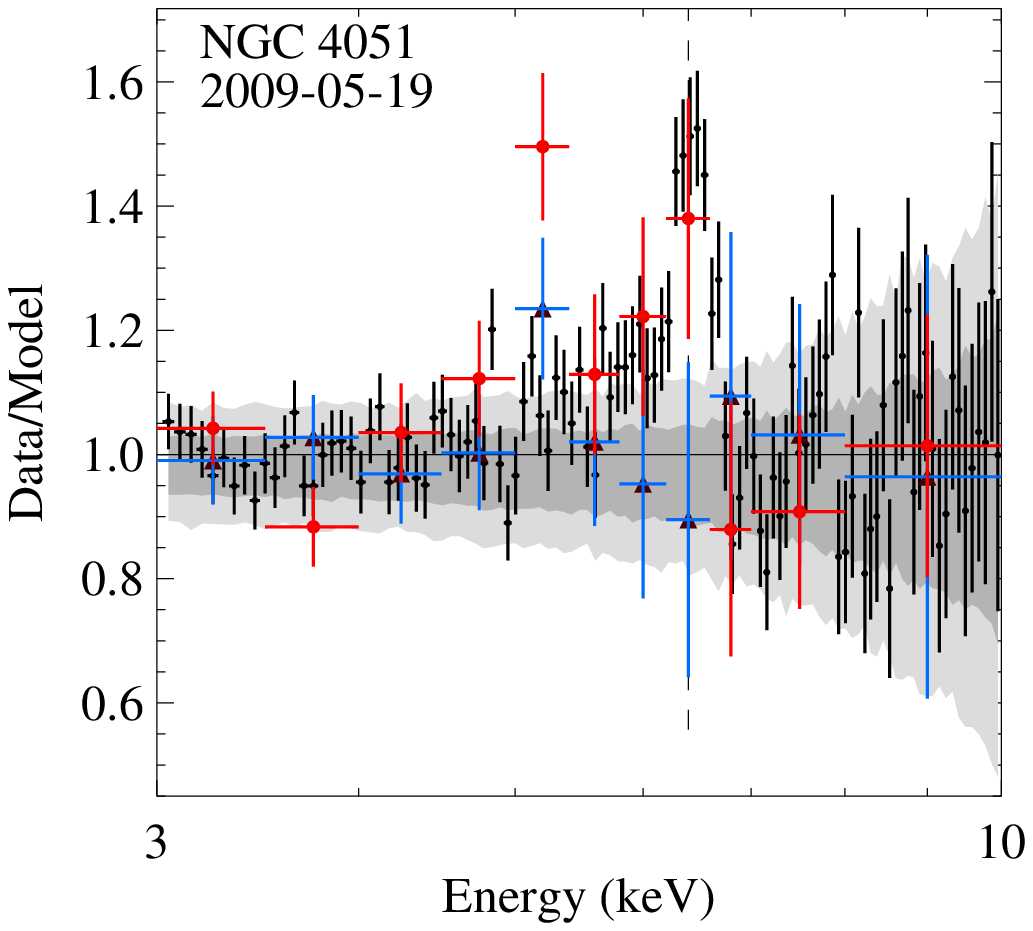} 
\includegraphics[width=1.75in]{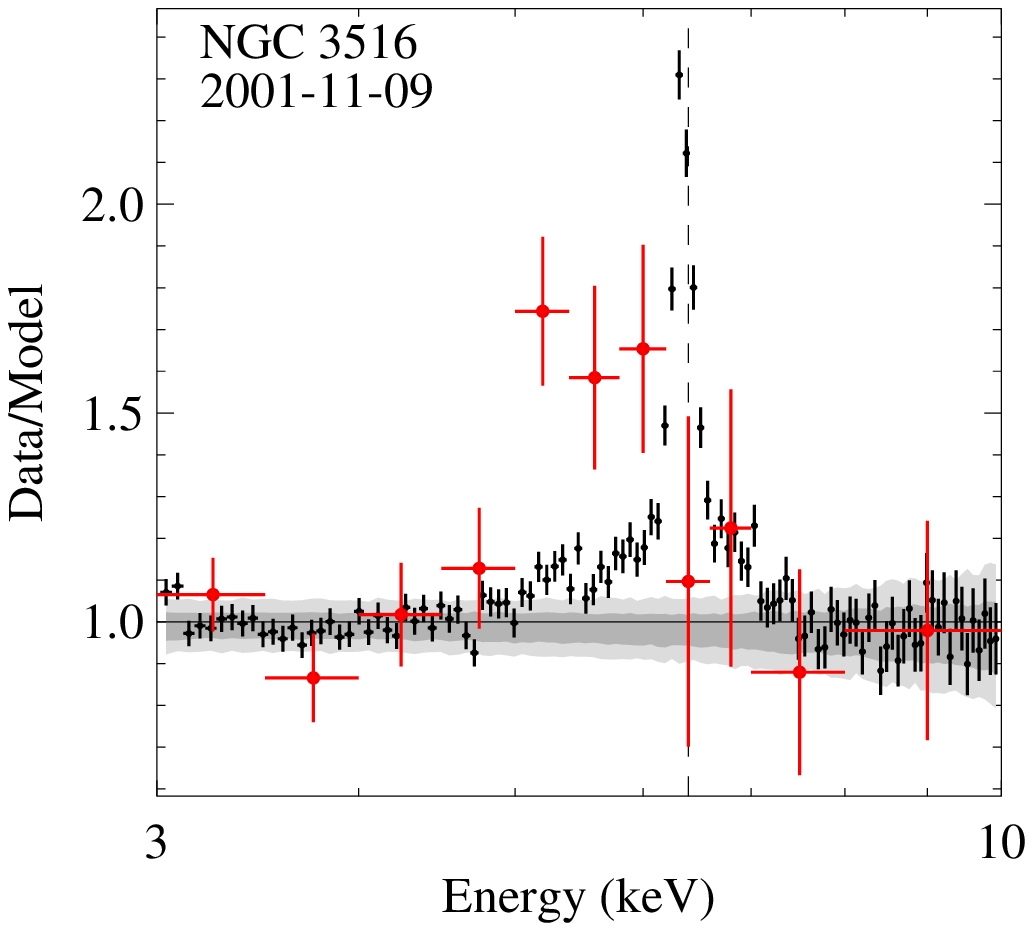} 
\caption{Some examples of the time-averaged spectra (black), LF rms spectra (red), and HF rms spectra (blue) around the Fe K$\alpha$ band. The 1$\sigma$ and 2$\sigma$ confidence intervals of the Fe line in the time-averaged spectra are marked as dark and light shaded areas, respectively. The first row shows some examples when a narrow or relatively symmetric Fe line is present, in which case no additional variability is observed. The second row shows some examples when a broad and redshifted Fe line is present in the time-averaged spectra, which is accompanied by additional LF and HF variability.}
\label{fig:Fe}
\end{figure*}

\section{Variability of the Fe K$\alpha$ Line} \label{FeK}
In many time-averaged spectra, we can visually identify a prominent X-ray Fe K$\alpha$ fluorescence line a rest-frame energy of 6.4 keV \citep{2007ARA&A..45..441M,2015ARA&A..53..365N}.
The narrow Fe K$\alpha$ line is often considered to be produced by reprocessing of the central X-ray coronal emission by distant materials such as the dusty torus \citep{1987ApJ...320L...5K,2006MNRAS.368L..62N}.
The broad Fe K$\alpha$ line is often considered to originate from the innermost part of the accretion disk, with its asymmetric profile caused by the relativistic beaming effect and gravitational redshift \citep{1989MNRAS.238..729F,2000PASP..112.1145F}.
Such a broad component of the Fe K$\alpha$ line has been detected in many AGNs \citep[e.g.][]{1997ApJ...477..602N,2007ARA&A..45..441M,2007MNRAS.382..194N,2012MNRAS.426.2522P,2015MNRAS.447..517L,2016MNRAS.463..684L,2019MNRAS.488.4378H}, many of which are also included in our sample. Given the large distance and scale of the reprocessor, it can be inferred that the narrow component may be much less variable than the broad component. Observationally, \citet{2010MNRAS.408.1020B} presented an rms variability analysis of a sample of 18 observations of 14 Seyfert galaxies observed by {\it XMM-Newton}. It is found that the narrow core of the K$\alpha$ line at 6.4 keV shows minimal evidence for variability and is always less variable than the continuum, while half of observations do show evidence for variations in the wider Fe K band. Thus the LF and HF rms spectra in our sample can be used to distinguish these line components.

\subsection{Measuring the Line Flux and Variability}
To study the Fe line, the first step is to obtain the best-fitting continuum for each of the objects in the sample.
The data below 3 keV are excluded due to the complexity of spectral components in this band.
The spectral data in the 5-7 keV range is also masked to avoid any contribution from the Fe K lines.
Then we fit the leftover spectra with a power-law model modified by Galactic and any possible intrinsic absorption, which is modeled with the \texttt{tbabs*ztbabs*zpowerlaw} model in \texttt{XSPEC} using the cross sections of \citet{1996ApJ...465..487V} and the interstellar medium abundances of \citet{2000ApJ...542..914W}.
The underlying continuum can be well fitted with a power law.
The excess emission features around 6.4 keV in the AGN rest frame appears to be the Fe K emission lines.

In order to distinguish the line variations from the continuum with sufficient resolution, we rebin the rms spectra and adopt the same set of 11 energy bins ranging from 3 to 10 keV for all the spectra.
The same power-law model is applied to these absolute rms spectra, as in the time-averaged spectra.
The photon index and normalization are left free to vary.
Then the variability around the Fe line region is investigated by deriving the ratio of the rms spectra to the best-fit power-law model.
The data-to-model ratios of several typical examples are plotted in Figure \ref{fig:Fe}.
The data points marked by the black crosses, red filled dots, and blue filled triangles with error bars represent the time-averaged spectra, and the absolute LF and HF rms spectra, respectively.

\subsection{Significance of the Line Flux and Variability } \label{significance_Fe}
We quantify the significance of the emission line and the line variability by means of Monte Carlo simulations. To examine the reliable detection of the Fe line, i.e. being statistically significant rather than due to random fluctuations, we simulate the `observed' X-ray spectra of the sample sources by assuming no Fe lines, following \citet[see also \citealt{2016MNRAS.463..684L,2016MNRAS.459.1602L,2019MNRAS.488.4378H}]{2015MNRAS.447..517L}.
The uncertainty in the best-fitting continuum model itself is also taken into account, following \citet[see also \citealt{2006MNRAS.366..115M,2010A&A...521A..57T}]{2006ApJ...646..783M}. The basic procedure is: for each of the sources in our sample, we first simulate a spectrum using \texttt{fakeit} by assuming the best-fitting continuum model, i.e. the zeroth-order null hypothesis. This fake spectrum is grouped to a minimum of 25 counts per energy bin for the next step. We fit again this grouped fake spectrum with the null hypothesis model and record the parameter values of the new model, i.e. a refined null hypothesis. Then we simulate a source and a background spectra using \texttt{fakeit} by assuming this refined null hypothesis model, which now includes counting statistics uncertainties. 
For each observation, the same exposure time, auxiliary and redistribution matrix file are used in these simulations to mimic the observed spectrum.
The above procedure is repeated 1000 times to produce 1000 simulated spectra for each observation.
We can estimate the 1$\sigma$ (68\%) and 2$\sigma$ (95\%) confidence intervals by calculating the 16.0-84.0th and 2.5-97.5th percentile values, respectively in each energy bin, which are shown as the dark and light shadowed areas in Figure \ref{fig:Fe}. It is clear that the emission feature around 6.4 keV is detected significantly in these time-averaged spectra.

To estimate the significance of the Fe line variability, a similar procedure is applied to the rms spectra, which are modeled with a power-law continuum with no emission line components.
For every data segment of 100 s bin, the same grouped fake spectrum with the refined null hypothesis model is adopted, which includes the uncertainties of the original best-fitting continuum model. The normalization of the refined null hypothesis model is rescaled so that the total count rate of the model matches the observed count rate in this data segment within 3-5 and 7-10 keV bands. 
Using this refined null hypothesis model and taking the random Poisson noise into account, we simulate a new fake spectrum, which provides fluxes in every energy bin. The above tasks are conducted for every data segment, which then allows us to produce fake light curves in every energy bin. Then we perform the same rms variability calculation.
Such a procedure is repeated 1000 times, so that the confidence level of rms spectra in each energy bin can be calculated by calculating how many times the simulated rms amplitude is in excess of the observed rms amplitude. An excess in one energy bin is marked as significant if its confidence level exceeds 95\%.

\subsection{The Diverse Fe Line Variability} \label{res_Fe}
An excess emission feature around 6.4 keV (more than three adjacent data bins in excess of the best-fitting continuum) is shown in the majority of the time-averaged spectra of our sample (71 sources with 359 observations, i.e. 91\% for source, 84\% for observation). When the selection criterion is chosen to be more than three adjacent data bins around 6.4 keV in excess of the 2$\sigma$ confidence region, the Fe line can be significantly found in $\sim$60\% of our sample (48 sources with 241 observations). This is broadly consistent with the result of \citet{1994MNRAS.268..405N}, who presented an analysis of 60 {\it Ginga} spectra of 27 Seyfert galaxies and found that Fe lines are detected at $>$99\% confidence in 37 of the 60 observations, and in 17 of the 27 objects.

Focusing on the Fe line variability, we consider objects with a significant spectral feature around the 6.4 keV region (more than three adjacent data bins in excess of the 2$\sigma$ confidence region) and fully constrained rms spectra only.
This results in 34 sources with a total of 140 {\it XMM-Newton} observations, including 140 LF rms spectra and 51 HF rms spectra.
Then the Fe line variability can be indicated by the significant excess of the rms spectra in this region over the fitted continuum.
The statistical results are listed in Table \ref{tab:Fe}.

\begin{deluxetable}{lcc}
\tablenum{4}
\tablecaption{Number of observations showing variability around the Fe K$\alpha$ line at the 95\% confidence level in certain energy bins \label{tab:Fe}}
\tablewidth{0pt}
\tablehead{
\colhead{Energy bands} & \colhead{LF rms variable} & \colhead{HF rms variable}
}
\startdata
5.0-6.2 keV        & 21  & 9    \\
6.2-6.6 keV       & 10 & 2    \\
6.6-7.0 keV        & 7  & 1    \\
\enddata
\tablecomments{The number of total LF and HF rms spectra in use are 140 and 51, respectively.}
\end{deluxetable}

The narrow Fe K$\alpha$ line is considered to be constant because of the long distance between the reflecting materials (e.g. the torus) and the primary X-ray emission close to the black hole. To check this, we explore the variability in the 6.2-6.6 keV energy band. We find that most observations showing a significant Fe K$\alpha$ line do not show significant LF or HF variability in this energy band (i.e. 49/51 HF rms spectra and 130/140 LF rms spectra show no variability), consistent with previous results \citep[e.g.][]{2010MNRAS.408.1020B}. Typical examples are shown in the first row of Figure \ref{fig:Fe}, where the line appears to be narrow, strong, and relatively symmetric, and shows no variability. This is consistent with the physical interpretation of the narrow-line component originating from the reflection of distant materials.

The broad Fe K$\alpha$ line is considered to originate from the inner region of the accretion disk, and its asymmetric profile arises from the relativistic effect close to the black hole. Thus this line component can be more variable than the narrow component. To test this, we explore the excess of the red and blue wings of the Fe K$\alpha$  line in the rms spectra. Indeed, we find that there are about 15\%-18\% of the sample (9/51 HF rms spectra and 21/140 LF rms spectra) show variability between 5.0-6.2 keV, and about 2\%-5\% of the sample (1/51 HF rms spectra and 7/140 LF rms spectra) show variability between 6.6-7.0 keV. Typical examples are shown in the second row of Figure \ref{fig:Fe}, where the Fe K$\alpha$ line profile are broad and asymmetric, generally showing an extended red wing. Despite the relatively low S/N, the rms spectra also exhibit an extended red wing. However, the rms line profile does not follow exactly the time-averaged line profile, indicating further complexities in the line components.

\section{Discussion} \label{discussion}

\subsection{Interpretations of Different Rms Spectra} \label{connectionDiscusstion}
A key result of this work is the discovery of the wide variety of rms spectra in a large sample of Seyferts with a large number of deep observations. The shapes of the rms spectra are found to be frequency-dependent. It should be noted that, as the PSD of AGN is dependent on the black hole mass \citep[e.g.][]{2006Natur.444..730M,2012AA...542A..83P,2015ApJ...808..163P,2018ApJ...858....2G}, the ideal boundary dividing the HF and LF bands may be different from one source to another. The objective of this work is to show the frequency dependence of the rms spectra, and as an approximation we simply chose $10^{-4}$ Hz as the boundary for all the sources.

Our results reveal that AGN do not only show different time-averaged spectra but also show different rms spectra. Based on previous spectral-timing studies of some individual AGN, we discussed possible interpretations to some specific combinations of the time-averaged spectra and rms spectra, which may help to understand other sources with similar spectral-timing properties.

Firstly, sources in the spec1 group may suffer from severe neutral absorption, but they also show different types of rms spectra. Sources showing both spec1 and rms2 can be explained by the emergence of a stable component in the soft X-ray band, which probably originates from the Compton scattering of distant materials such as the torus \citep[e.g.][]{2014ApJ...791...81A}. Another possibility is a stable warm Comptonization component as observed in the soft excess of the NLS1 RE J1034+396 \citep[e.g.][]{2011MNRAS.417..250M,2021MNRAS.500.2475J}, but this component may suffer from severely absorption as expected in spec1.
In addition, it may also be due to the emergence of large-scale X-ray emission such as photoionized gas \citep[e.g.][]{2021MNRAS.500.4506H}, scattering, and/or star-formation, as seen in heavily absorbed (e.g. Compton thick) AGN \citep[e.g.][]{2012ApJ...758...82L,2019ApJ...887..173L}.
Sources showing spec1, HF rms4, and LF rms3 can be understood if the absorbing materials introduce variability at low frequencies by changing, e.g. the covering factor \citep{2021MNRAS.508.1798P}.

Secondly, sources in the spec2 group tend to show rms2. This combination of spectral-timing properties is qualitatively consistent with the ionized reflection scenario. In this case, the ionized reflection component creates a flux dip at 1-2 keV, which leads to the spec2 shape \citep[e.g][]{2011MNRAS.418.2642Z,2013MNRAS.428.2795K,2019MNRAS.489.3436J}. Then as the underlying continuum from the Comptonization of the hot corona contributes more variability, an rms peak emerges at the same energy band which belongs to the rms2 group \citep{2020MNRAS.492.1363P}. Another scenario is the ionized (warm) absorber \citep[e.g.][]{2003ARA&A..41..117C,2004MNRAS.353..319T}, in which case the ionized absorption creates a flux dip at 1-2 keV, and the variation in the ionization state and/or covering fraction introduces additional variability \citep{2021MNRAS.508.1798P}.

Thirdly, sources in the spec3 group do not show any dip in the flux between 1 and 2 keV, the most common rms for spec3 is rms2, as for spec2. It may due to a variable continuum damped with a less-variable soft excess, more components such as relativistic reflection and/or ultrafast outflow are also included in some cases \citep[e.g.][]{2021MNRAS.508.1798P}. In addition to the rms2, 
many sources in the spec3 group show LF rms3 and HF rms1. As mentioned earlier, these properties are typical of some ``X-ray simple" super-Eddington NLS1s, such as PG 1244+026 and RX J0439.6-5311, whose X-ray spectra are both steep and smooth, and show a significant soft excess. One physical model to reproduce this type of time-averaged spectrum is the ionized reflection. The smoothness of the spectrum requires the reflection emission to be smeared by the strong relativistic effect, and so an extreme black hole spin is often required \citep[e.g.][]{2011MNRAS.418.2642Z,2014MNRAS.439L..26K}. However, the simple reflection model is difficult to reproduce the frequency-differentiated rms spectra of these sources \citep[e.g.][]{2013MNRAS.436.3173J,2017MNRAS.468.3663J}. Another scenario to explain this combination of spectral-timing properties is the warm corona model, where the soft excess is produced by the Comptonization of an extended warm corona, which is distinctive from the compact hard X-ray corona. The different sizes of these two regions lead to different variability timescales, which can explain the stronger LF variability in the soft excess and stronger HF variability in the hard X-rays. It can also explain the LF time lag observed in these sources \citep[e.g.][]{2013MNRAS.436.3173J,2017MNRAS.468.3663J,2021MNRAS.500.2475J}.

\begin{figure*}[ht!]
\includegraphics[width=6.8in]{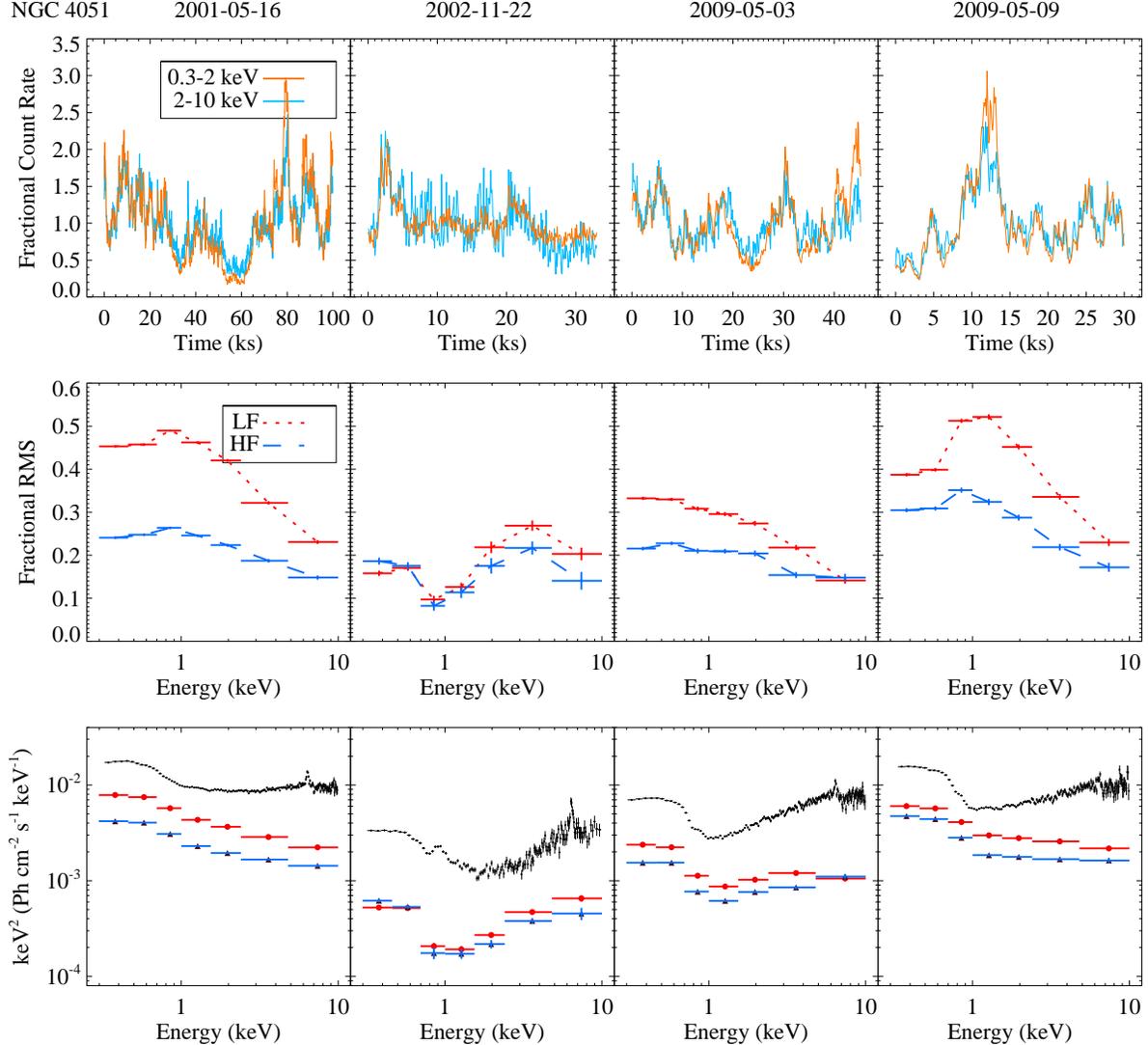}
\caption{Example of the change in the rms spectral over time for the case of NGC 4051. The top panel shows the fractional count rate variability of 0.3-2 keV and 2-10 keV, shown as orange lines and light blue lines, respectively. The middle panel shows the fractional rms in different frequency ranges. The LF ($<10^{-4}$ Hz) and the HF ($10^{-4}$ - $10^{-3}$ Hz) are shown in red dotted lines and blue dashed lines, respectively. The bottom panel shows the time-averaged spectrum (black crosses) and absolute rms spectra in the case of LF (red filled dots) and HF (blue filled triangle).}
\label{fig:change_rms}
\end{figure*}

As a more quantitative discussion in the framework of the above model, we assume that the X-ray variations may originate from thermal fluctuations in the heating and/or cooling of the gas in the warm and hot corona. The thermal timescale (i.e. the time needed for readjustment to thermal equilibrium) associated with the radiation region can be expressed as,
\begin{equation}
    t_{\rm thermal}\sim\frac{1}{\alpha\Omega}\approx1.4\times10^{-5}~\alpha^{-1}M~r^{\frac{3}{2}}k_{\Omega}^{-1}
\end{equation}
\citep{1992apa..book.....F, 2008bhad.book.....K}, in which $\alpha$ is the viscosity parameter, $M$ is the black hole mass in units of $M_{\odot}$, $r$ is the radius in units of Schwarzschild radius $R_{\rm S}=2GM/c^2$ and $k_{\Omega}$ is the ratio of angular velocity $\Omega$ to the Keplerian velocity $\Omega_{\rm K}$ ($k_{\Omega}=1$ for the standard Keplerian disk). As a rough estimation, $\alpha$ is set to a typical value of $0.3$ and $M$ to the mean value of spec3 subsample of $10^{7}~M_{\odot}$. For the optically thin hot corona, we employ the solution of the advection-dominated accretion flow (ADAF) of $k_{\Omega}\approx0.4$ \citep{2014ARA&A..52..529Y}, then a typical coronal size of a few Schwarzschild radius ($r\sim3-5$) measured from reverberation mapping and gravitational lensing \citep[e.g.][]{2013ApJ...769L...7R, 2015MNRAS.451.4375F, 2016AN....337..356C} leads to $t_{\rm thermal}\approx6.1\times10^3 - 1.3\times10^4$~{\rm s}, which is broadly consistent with the HF variability timescale ($\sim10^{3}-10^{4}~{\rm s}$). For the warm corona, we take $k_{\Omega}=1$ for its more disk-like geometry. Also, the warm corona is generally more extended in the radial direction, spanning from $\sim10R_{\rm S}$ to several tens of $R_{\rm S}$ \citep{2017MNRAS.468.3663J}. Therefore, for $r$ ranging from 10 to 50, $t_{\rm thermal}\approx1.5\times10^{4} - 1.6\times10^{5}~{\rm s}$, which may also explain the LF variability ($>10^{4}~{\rm s}$). 

We note that the HF variability associated with the hot corona could also be triggered by density fluctuations. The viscous timescale (i.e. the time over which viscosity smooths out surface density gradients) can be expressed as,
\begin{equation}
    t_{\rm viscous}=t_{\rm thermal}(\frac{R}{h})^2\sim\frac{1}{\alpha\Omega}(\frac{R}{h})^2
\end{equation}
in which $h$ is the vertical scale-height of the accretion flow. For the geometrically thick ($h\approx R$), ADAF-like hot corona, the viscous timescale is approximately equal to the thermal timescale. The case for the warm corona, however, remains more uncertain due to the poor knowledge of its vertical structure. By comparing the radius of the soft X-ray region and that of the geometrically thick inner disk region dominated by the radiation pressure, \citet{2017MNRAS.468.3663J} found tentative evidence for the connection between the warm corona and the puffed-up inner disk, hinting that we may take the scale-height of the inner disk as a reasonable approximation. Specifically, for a radiation pressure dominated Shakura-Sunyaev disk spanning from $\sim10R_{\rm S}{\rm to}50R_{\rm S}$, the $R/h$ ratio is around $\sim2-6$ at a critical accretion rate $\dot{m}\approx1$ ($\dot{m}=\dot{M}/\dot{M}_{\rm Edd}$, $\dot{M}_{\rm Edd}=3\times10^{-8}~m~\frac{M_{\odot}}{\rm yr}$ is the Eddington accretion rate for a Schwarzschild black hole). Consequently, the viscous timescale is expected to be larger than the thermal timescale by a factor of several to several tens, increasing rapidly with the distance to the central black hole. Therefore, we suggest that the density fluctuations in the warm corona may be responsible for the variability at a longer timescale than those concerned in this work, except for those emerging at innermost regions, or in some super-accreting systems with very high accretion rates \citep[the scale-height depends linearly on the accretion rate for a radiation pressure dominated disk, {$h\propto\dot{m}$},][]{1973A&A....24..337S}. 
It should be noted that the above calculations are only order-of-magnitude estimates and the obtained values are subject to large uncertainties.

The above interpretations are simply based on previous studies of a few individual sources. As different combinations of time-averaged spectra and rms spectra may correspond to different physical mechanisms, further studies are required in order to interpret other combinations.

\subsection{Variation of the Rms in Individual Sources} \label{DespersionDiscusstion}
Recent studies on individual sources have shown that the rms spectrum and power spectrum can change significantly as the time-averaged spectrum changes between observations \citep[e.g.][]{2021MNRAS.508.1798P,2022arXiv220712151Y}. As shown in Table \ref{tab:sample}, the majority of our sample have multiple observations, which allows us to investigate the variability of the rms spectra in these sources.

Figure \ref{fig:change_rms} shows NGC 4051 as a typical example. Its rms spectra change significantly as the time-averaged spectra change. During the first {\it XMM-Newton} observation in 2001 May 16, both fractional LF and HF rms spectra display a concave-down shape peaking at 1-2 keV. Then during the observation in 2002 November 22, the flux of the time-averaged spectra decreases by almost 1 order of magnitude, and the fractional rms in both LF and HF bands are strongly suppressed below 3 keV. During the third observation in 2009 May 03, the flux recovers by a factor of few, and the rms in LF and HF are also observed to increase. During the fourth observation in 2009 May 09, the time-averaged spectrum and the LF and HF rms spectra all return to a similar state as the first observation. This example clearly reveals the connection of the variation between the rms and time-averaged spectra in individual sources.
A possible physical interpretation is that the photoionized gas emission on large scale dominates the soft X-ray band at low fluxes, as at high fluxes the continuum dominates and the contribution from photoionized gas is negligible, similar to the recent study on 1H 0707-495 by \citet{2021MNRAS.508.1798P}. In addition, it may also be due to the variability of the ionized absorption along the line of sight. However, the Eddington ratio of NGC 4051 is quite low, so the absorber may have different properties. Based on the study of our sample, we point out that the covariation in the rms and time-averaged spectra is common, as shown in Figure \ref{src_rmsspec} in the appendix, but a more detailed study of individual sources is beyond the scope of this work.

\subsection{Origin of the Fe K$\alpha$ Line} \label{FeKDiscusstion}
About 60\% of the time-averaged spectra of our sample exhibit a significant spectral feature around 6.4 keV, for which the rms spectra with small energy bins around this region are produced.
The majority of the Fe line shows no significant variability in the HF and LF rms spectra, partly due to the relatively low count rate in this energy band. Interestingly, we do detect significant variability in 140 observations of 34 sources.

The narrow Fe K$\alpha$ line is commonly present in AGN spectra and is thought to be the fluorescence emission from distant materials such as the dusty torus. It is not expected to show strong variability, which is consistent with our results. The broad Fe K$\alpha$ line is believed to be produced by the reflection of corona emission in the inner disk region \citep[e.g.][]{2013MNRAS.428.2795K,2013MNRAS.429.2917F,2015MNRAS.449..234K}. In this scenario, the reflection should respond rapidly to the change in the corona, so that significant variability is expected in the broad Fe K$\alpha$ line. Indeed, significant time lags have been reported for a sample of Seyferts \citep{2013MNRAS.431.2441D,2016MNRAS.462..511K}.

Despite the relatively low S/N, we notice that the rms spectra of the Fe K$\alpha$ line do not show similar profiles as the time-averaged spectra; this suggests that the Fe K$\alpha$ line profile may contain various components of different variability properties, similar to the properties of hydrogen Balmer lines in the optical \citep[e.g.][]{2004ApJ...613..682P}. Next generation telescopes such as {\it Athena} \citep{2013arXiv1306.2307N} and {\it eXTP} \citep{2016SPIE.9905E..1QZ} will be able to provide data with much higher S/N to reveal more detailed properties of the Fe K$\alpha$ line.

\section{Summary and Conclusion} \label{conclusion}
We have conducted a systematic study of the X-ray variability for a large sample of 78 Seyfert galaxies observed with 426 deep {\it XMM-Newton} observations. In this paper, we focus on the rms spectra in both low and high-frequency bands. The main results are summarized below.

\begin{itemize}
\item[1.] We find that the calculation of the intrinsic rms is sensitive to the subtraction of the Poisson noise power. The adoption of theoretical Poisson noise power can lead to significant underestimate of the true Poisson noise due to various corrections applied to the observed light curve. Thus the intrinsic rms can be overestimated. Although it is possible to apply further corrections, we suggest that the most accurate method is to calculate the arithmetic average of the power in the HF band where the Poisson noise power dominates.

\item[2.] We find a wide variety of rms spectra and time-averaged spectra in the sample. Our statistical study of the sample confirms the correlation between the Eddington ratio and the steepness of the time-averaged spectra. We also find that sources with the largest black hole masses and low Eddington ratios in our sample tend to show both flat rms spectra and time-averaged spectra.

\item[3.] There exist various combinations of the time-averaged spectra and rms spectra, i.e. there is no one-to-one mapping between different subtypes of the time-averaged spectra and rms spectra. The same time-averaged spectra can have different rms spectra, while the same rms spectra can be accompanied by different time-averaged spectra.

\item[4.] The most common rms spectral shape is the one which has a concave-down shape with an rms peak at $\sim$1 keV (rms2). The rarest rms shape is the one which has a concave-up shape with an rms minimal at $\sim$1 keV (rms5). The shape of rms2 is qualitatively consistent with the rms pattern introduced by the ionized reflection or the ionized absorption, while the shape of rms5 is qualitatively consistent with the rms pattern introduced by the photoionized gas emission on large scale.

\item[5.] Comparing the time-averaged spectral groups of spec1, 2, and 3, we find that they all prefer the rms2 spectral shape, but there are also systematic differences among the three spectral groups in terms of both observed properties and theoretical interpretations.
For spec1, the HF rms spectra tend to be flat, and the LF rms spectra tend to show more fractional rms towards soft X-rays. This indicates that neutral absorption cannot introduce additional HF variability but may introduce LF variability.
For spec2, the spectral-timing properties can be interpreted as a variable power law damped with a less-variable soft excess which may (or may not) be associated with relativistic reflection and/or a warm corona.
For spec3 with the highest Eddington ratio, the soft X-rays tend to show weaker HF rms, while the hard X-rays tend to show stronger HF rms. These properties are similar to those found in super-Eddington NLS1s and are consistent with the warm Comptonization model for the soft excess.

\item[6.] Significant variability is detected at the Fe K$\alpha$ line energies in 34 Seyferts with 140 observations. We find that broad Fe K$\alpha$ lines with a redshifted-wing tend to show significant variability, which is likely because the line originates from disk reflection close to the black hole. The narrow Fe K$\alpha$ line with a symmetric line profile is much less variable, which is consistent with the distant reflection model.

\end{itemize}

Our results show that the detailed variability analysis can provide an important diagnostics, complementary to the time-averaged spectral analysis. Thus a joint spectral-timing analysis is valuable for the understanding of the X-ray radiation mechanisms of AGN. To make such a study easier for the community and to make the most of the valuable {\it XMM-Newton} observational data, we have made all the time-averaged spectra and HF/LF rms spectra of the entire sample publicly available on our website\footnote{\url{https://gohujingwei.github.io/rms.html}. All the spectral files can be downloaded directly for joint modeling in {\sc xspec}.}.

\acknowledgments
We thank the anonymous referee providing valuable comments and suggestions, which have improved the quality of the paper.
C.J. acknowledges Wenda Zhang for valuable discussions.
This work is supported by the National Natural Science Foundation of China (grant Nos.11773037, 11673026, 11473035, 11873054, 11803047, 12173048), the gravitational wave pilot B (grant No.XDB23040100), the Strategic Pioneer Program on Space Science, Chinese Academy of Sciences, grant No.XDA15052100.
This work is based on observations conducted by {\it XMM-Newton}, an ESA
science mission with instruments and contributions directly funded by
ESA Member States and the USA (NASA).

\bibliography{rms}{}
\bibliographystyle{aasjournal}

\appendix

\section{The Frequency-resolved Rms Spectra} \label{rms}
\label{sec-rms}

\subsection{Calculation of the Rms Spectra}
\label{sec-calRms}

The fractional rms amplitude \citep{2002ApJ...568..610E,2003ApJ...598..935M,2003MNRAS.345.1271V} is a common measure of the intensity of intrinsic source variability. In this work, we adopt the frequency-resolved spectral techniques developed by \citet{1999A&A...347L..23R} to calculate the rms from X-ray light curves. We summarize the calculation procedure below \citep[see also][]{2003MNRAS.345.1271V, 2008MNRAS.389.1427P,2017MNRAS.468.3663J}. A detailed prescription can also be found in \citet{2008MNRAS.387..279A}. 

The PSD quantifies the amount of variability power as a function of frequency. A periodogram, which is a single realization of a PSD, can be obtained through the discrete Fourier transform \citep[DFT;][]{1992nrfa.book.....P} of a light curve as,

\begin{equation}
\label{equ1}
P(f_i)=\frac{2\Delta{T}}{N \bar{x}^2} \left| {\rm DFT} (f_i) \right|^2=\frac{2\Delta{T}}{N \bar{x}^2} \left| \sum_{j=1}^N x_j e^{2\pi i f_i t_j} \right|^2
\end{equation}
where the power density $P(f_i)$ is in units of $\rm Hz^{-1}$, $\Delta{T}$ is the time bin of the light curve, $N$ is the number of useful time bins, $\bar{x}$ is the average count rate of the light curve $x(t_j)$, $f_i$ is the Fourier frequency defined as $f_i=i/T$. Each light curve is binned in 100 s and thus gives a Nyquist frequency (i.e. the upper limit of the frequency range of a periodogram) of $5\times10^{-3}$ Hz. Moreover, we divide the frequency range into two subranges, and define $10^{-4}$ Hz $<f \le 10^{-3}$ Hz as the HF band and $f \le 10^{-4}$ Hz as the LF band.

As we adopt the Belloni-Hasinger normalization \citep{1990A&A...227L..33B} in Equation~\ref{equ1}, the normalized excess variance $\sigma^2_{\rm NXS}$ can be calculated by integrating the periodogram over a specific frequency range after subtracting the Poisson noise power \citep{2008MNRAS.387..279A},

\begin{equation}
\sigma^2_{\rm NXS,\Delta{f}}=\sum_{i=a}^b (P(f_i)-\rm PN_{lev})\delta{f}
\end{equation}
where $\Delta{f}$ is the frequency range of $\Delta{f}=b/T-a/T$, $\delta{f}$ is defined as $\delta{f}=1/T$, $\rm PN_{lev}$ is the Poisson noise contribution calculated by averaging the power density in the frequency range of larger than $2\times10^{-3}$ Hz, where the Poisson noise power dominates. Full details of the Poisson noise calculation are given in Section \ref{PoissonNoise}.
Then the fractional rms $\sigma_{{\rm rms},\Delta{f}}$ can be calculated as,

\begin{equation}
\label{equ2}
\sigma_{rms,\Delta{f}}=\sqrt{\sum_{i=a}^b (P(f_i)-\rm PN_{lev})\delta{f}}
\end{equation}

The error of $\sigma_{\rm rms}$ can be calculated using the following equation \citep{2008MNRAS.389.1427P},

\begin{equation}
\label{equ3}
\begin{aligned}
& {\rm err}(\sigma_{{\rm rms},\Delta{f}})= 
& \sqrt{ \sigma^2_{{\rm NXS},\Delta{f}} + {\rm err}(\sigma^2_{{\rm NXS},\Delta{f}}) }-\sigma_{{\rm rms},\Delta{f}}
\end{aligned}
\end{equation}
where ${\rm err}(\sigma^2_{\rm NXS,\Delta{f}})$ can be calculated as, 

\begin{equation}
\begin{aligned}
& {\rm err}(\sigma^2_{\rm NXS,\Delta{f}})= 
& \sqrt{\left(\frac{{\rm PN_{lev}}\times\Delta f}{\sqrt {N'}}\right)^2 + \frac{{\rm 2PN_{lev}}\times\Delta f \sigma^2_{\rm NXS,\Delta{f}}}{N'}}
\end{aligned}
\end{equation}
where $N'$ is the number of frequency points within $\Delta{f}$ in the periodogram.

Then the absolute rms spectra can be calculated by multiplying the fractional rms by the average count rate in each energy bin.
The resultant absolute rms spectra can be fitted in \texttt{XSPEC} using the same response and ancillary files of the time-averaged spectra \citep[e.g.][]{2006A&A...447..545R,2006MNRAS.370..405S,2011MNRAS.417..250M,2013MNRAS.436.3173J}.
The variable component of a time-averaged spectrum in the LF and HF bands can be revealed directly by the corresponding absolute rms spectra. This means that if a time-averaged spectrum comprises several components with different variability properties in different frequency bands, then a joint modeling of the time-averaged spectra and absolute rms spectra can help to distinguish these components \citep[e.g.][]{2013MNRAS.436.3173J,2016MNRAS.455..691J,2017MNRAS.468.3663J,2020MNRAS.492.1363P,2020MNRAS.495.3538J,2021MNRAS.508.1798P}.

\subsection{Determination of the Poisson Noise Power}
\label{PoissonNoise}
A PSD comprises the power of both the intrinsic source variability and the Poisson noise, and so the subtraction of Poisson noise power can affect the accuracy of the intrinsic rms of the source. This is especially important when the signal-to-noise of the light curve is low, either due to the low source flux or the narrow energy band. To this end, we investigate the accuracy of various methods for the estimation of Poisson noise power, and test their impact on the resultant rms.

The Poisson noise power is supposed to be a constant component in the PSD at a theoretical level of ${\rm PN_{lev}}=2/\bar{x}$, where $\bar{x}$ is the average source count rate. If we also consider the background subtraction, then ${\rm PN_{lev}}$ can be calculated as \citep{2008MNRAS.387..279A},

\begin{equation}
\label{eq-pn}
{\rm PN_{lev}}=\frac{2(\bar{x}+\bar{b})}{\bar{x}^2}
\end{equation}
where $\bar{b}$ is the average count rate of the background underneath the source extraction region. It must be noted that the background light curve also contains Poisson noise power; thus during the background subtraction, the noise of background will also propagate into the final light curve. This means that in reality Equation~\ref{eq-pn} underestimates the Poisson noise power, unless the background can be perfectly determined.

However, our source light curves are produced by the \texttt{epiclccorr} task in the \texttt{SAS} software, which performs absolute corrections, relative corrections that are time-dependent, and background subtraction\footnote{https://heasarc.gsfc.nasa.gov/docs/xmm/sas/help/epiclccorr/index.html}. For example, our source extraction region only covers a fraction of the PSF; thus a correction factor larger than unity will be applied to the source light curve in order to recover the intrinsic source flux. Therefore, the theoretical ${\rm PN_{lev}}$ calculated from Equation~\ref{eq-pn} for a corrected light curve will be smaller than the true ${\rm PN_{lev}}$ contained in the light curve, and then the resultant fractional rms will be biased to a larger value. We demonstrate this effect in Figure~\ref{fig:psd}, which shows the periodograms of two sources. The cyan solid line indicates the Poisson noise power calculated from Equation~\ref{eq-pn}, which is clearly smaller than the constant white-noise power observed at high frequencies. We also note that this effect was never mentioned in previous literatures, and so we urge caution to be exercised in dealing with this effect.

\setcounter{figure}{0}
\renewcommand{\thefigure}{A\arabic{figure}}
\begin{figure}[ht!]
\centering
\includegraphics[width=2.7in]{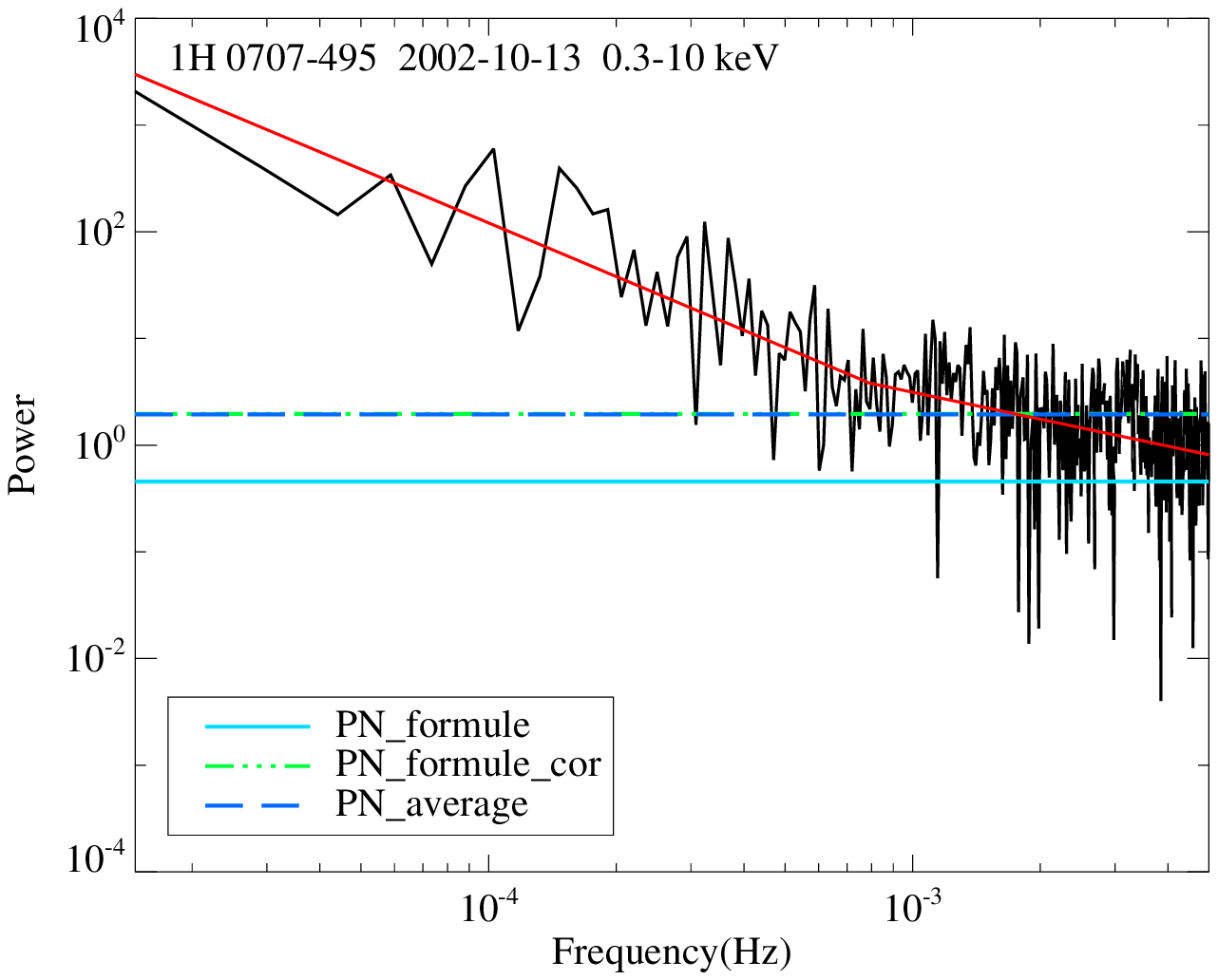}
\includegraphics[width=2.7in]{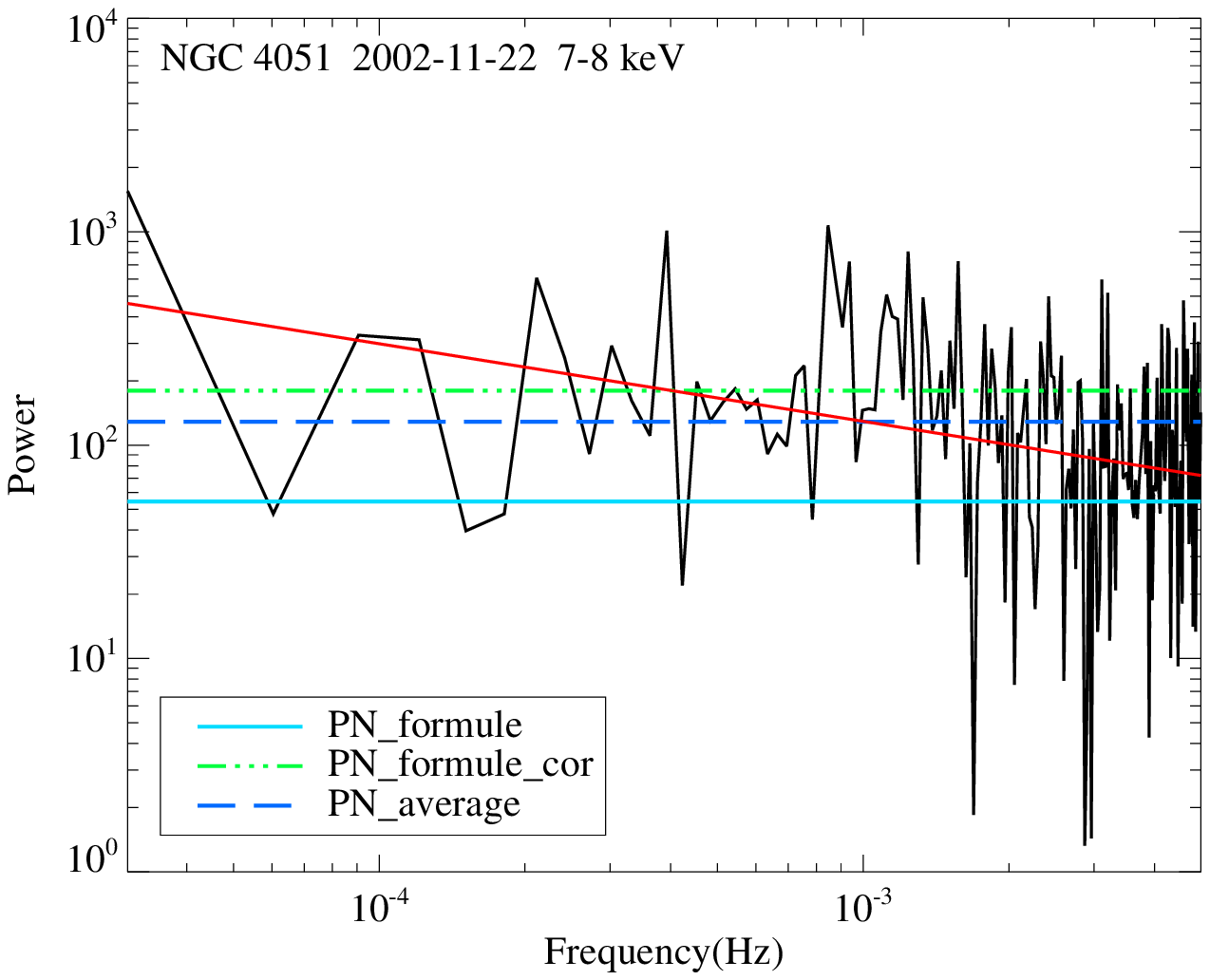}
\caption{Two examples of the PSD results. The best-fitted broken power laws are shown in red solid lines. The other horizontal lines mark the Poisson noise calculated from various method. The cyan solid lines represent the common formula, ${\rm PN_{lev}}=2(\bar{x}+\bar{b})/\bar{x}^2$, where $\bar{b}$ is the average count rate of the background, and $\bar{x}$ is the average count rate of the background-subtracted light curve. The green dotted–dotted–dotted–dashed lines represent the correction of this formula, ${\rm PN_{lev\_cor}}=2((\bar{x}/\bar{f})+\bar{b})/{(\bar{x}/\bar{f})}^2$, where $\bar{f}$ is the mean PSF correction factor. The blue dashed lines represent the arithmetic average of the power over the PSD frequency range of above $2\times10^{-3}$ Hz.}
\label{fig:psd}
\end{figure}

In order to recover the true ${\rm PN_{lev}}$, we can replace $\bar{x}$ with $\bar{x}/f$ in Equation~\ref{eq-pn}, where $f$ is the correction factor applied to the background-subtracted source light curve. However, it is difficult to know an accurate $f$ because it is calculated by \texttt{epiclccorr} internally and varies with time. As \texttt{epiclccorr} should have propagated errors properly and applied the same correction factor to the count rates and errors, we can get an estimate of $f$ in every time bin with the following equation,

\begin{equation}
f=err(x(t))/\sqrt{\frac{s(t)+b(t)}{\Delta t}}
\end{equation}
where $x(t)$ is the flux-corrected light curve produced by the \texttt{epiclccorr} task, $s(t)$ the source light curve before background subtraction, and $b(t)$ the background light curve {\it rescaled} to the size of the source extraction region. Then we can derive a mean correction factor $\bar{f}$, and use $\bar{x}/\bar{f}$ to replace $\bar{x}$ in Equation~\ref{eq-pn} to get a better estimate of the true ${\rm PN_{lev}}$. The green line in Figure \ref{fig:psd} shows the new estimate of ${\rm PN_{lev}}$, which is higher and more consistent with the constant white-noise power observed at high frequencies.

Another method is to measure ${\rm PN_{lev}}$ from the periodogram directly. \citet{1995A&A...300..707T} showed that the probability distribution of a periodogram at a specific frequency follows a $\chi^2$ distribution with two degrees of freedom; thus ${\rm PN_{lev}}$ can be estimated by taking the arithmetic average of the power over the frequency range where the Poisson noise power dominates. After visually checking all the periodograms, we choose the frequency range above $2\times10^{-3}$ Hz to measure ${\rm PN_{lev}}$. The blue dash line in Figure \ref{fig:psd} shows the averaged Poisson noise power, which is most consistent with the constant white-noise power observed at high frequencies. Therefore, we adopt this method to determine the true ${\rm PN_{lev}}$, which is then used to calculate the intrinsic rms variability. It should be noted that the above method requires a proper time bin of the light curve. The binning time is mainly determined by the following two factors. Firstly, a short time bin should be used to ensure that the periodogram can show the constant Poisson noise power at high frequencies. Secondly, the time bin should not be too short, as otherwise it will create many zero-count bins in the light curve, which will bias the periodogram at high frequencies. We find that for our sample a bin size of 50-100 s is appropriate for most observations, but larger time bins must be used for very faint sources.

\section{Typical results of our sample}

Here we plot some typical results of 78 Seyfert galaxies in Figure \ref{src_rmsspec} \footnote{All the spectral files and images can be downloaded from our website. \url{https://gohujingwei.github.io/rms.html}. }.

\setcounter{figure}{1}
\renewcommand{\thefigure}{A\arabic{figure}}
\begin{figure*}[ht!]
\centering
\includegraphics[width=6.4in]{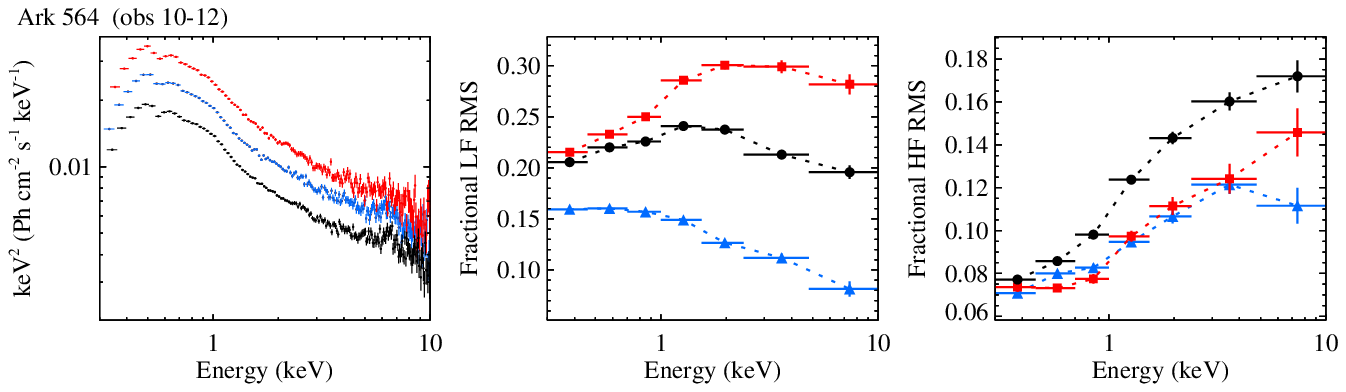}
\caption{Connection of the rms spectrum with the time-averaged spectrum for the sample. The complete figure set (78 images) is available in the online journal. Each three observations of one source are plotted in one picture, and different colors are defined as flux states. The lowest, intermediate, and highest fluxes are shown in black, blue, and red, respectively. Displayed in black if there is only one observation, in black and red if there are two observations, but always with black as representative of the low flux and red of the high flux. The left panel shows the time-averaged energy spectra for each of the individual observations used in this source. The middle panel shows the fractional rms in LF ($<10^{-4}$ Hz). The right panel shows the fractional rms in HF ($10^{-4}$ - $10^{-3}$ Hz).}
\label{src_rmsspec}
\end{figure*}

\end{document}